\documentclass[10pt,A4paper,twocolumn,aps,pra, superscriptaddress,longbibliography, nofootinbib]{revtex4-2}

\usepackage{amsmath}    
\usepackage{graphicx}   
\usepackage{color}      
\usepackage[usenames,dvipsnames,svgnames,table]{xcolor}
\usepackage[hidelinks]{hyperref}   
\usepackage{braket}
\usepackage{bbm}

\usepackage[capitalise]{cleveref}
\Crefname{figure}{Fig.}{Figs.}
\Crefname{table}{Tab.}{Tabs.}
\usepackage{siunitx}
\usepackage{amssymb}

\usepackage{rotating}
\usepackage{tabularx}
\usepackage{cellspace}
\usepackage{booktabs}
\usepackage{graphics}
\usepackage{newfloat}
\newcolumntype{Y}{>{\centering\arraybackslash}X}
\newcolumntype{d}{c}
\newcolumntype{a}{c}

\usepackage{savesym}
\savesymbol{printendnotes}
\usepackage{enotez}
\restoresymbol{NT}{printendnotes}
\newcommand{\emptymark}[1]{}
\setenotez{list-name=Methods, mark-cs=\emptymark}

\newcommand{\trace}{\ensuremath{\mathrm{Tr}}}
\newcommand{\natmethods}[1]{{ #1}}

\newcommand{\CZ}{CZ gate}
\newcommand{\CZs}{CZ gates}

\newcommand{\lz}{\ensuremath{\ket{0}_{\mathrm{L}}}}
\newcommand{\lo}{\ensuremath{\ket{1}_{\mathrm{L}}}}
\newcommand{\lp}{\ensuremath{\ket{+}_{\mathrm{L}}}}
\newcommand{\lm}{\ensuremath{\ket{-}_{\mathrm{L}}}}
\newcommand{\lopz}{\ensuremath{\hat{Z}_{\mathrm{L}}}}
\newcommand{\lopx}{\ensuremath{\hat{X}_{\mathrm{L}}}}
\newcommand{\px}{\ensuremath{\hat{X}}}
\newcommand{\pz}{\ensuremath{\hat{Z}}}
\newcommand{\sx}[1]{\ensuremath{\hat{S}^{\textrm{X}#1}}}
\newcommand{\sz}[1]{\ensuremath{\hat{S}^{\textrm{Z}#1}}}
\newcommand{\sai}{\ensuremath{\hat{S}^{Ai}}}

\newcommand{\ns}[1]{\ensuremath{\SI{#1}{\nano\second}}}
\newcommand{\midqbm}{\ensuremath{Qi}}
\newcommand{\midqbd}{\ensuremath{Qj}}

\newcommand{\fcdeltaphase}{\ensuremath{\phi_j}}
\newcommand{\fcdeltafreq}{\ensuremath{\omega_j}}
\newcommand{\fcdeltaflux}{\ensuremath{\Phi_j}}
\newcommand{\fcmtx}{\ensuremath{\mathbf{C}}}
\newcommand{\fcorrectable}{\ensuremath{\mathcal{F_\mathrm{c}}}}
\newcommand{\prlogicalsubspace}{\ensuremath{\hat{\mathcal{P}}_\mathrm{L}}}
\newcommand{\prlogicaloperator}{\ensuremath{\hat{\mathcal{P}}_\mathrm{O}}}

\newcommand{\puttitle}{Realizing Repeated Quantum Error Correction in a Distance-Three Surface Code}

\definecolor{adpcolor}{rgb}{0.36,0.54,0.66}

\begin{document}

\title{\puttitle}
\author{Sebastian Krinner}
\thanks{These authors contributed equally to this work.}

\author{Nathan Lacroix}
\thanks{These authors contributed equally to this work.}
\affiliation{Department of Physics, ETH Zurich, CH-8093 Zurich, Switzerland}

\author{Ants Remm}
\affiliation{Department of Physics, ETH Zurich, CH-8093 Zurich, Switzerland}

\author{Agustin Di Paolo}
\affiliation{Institut Quantique and D\'{e}partement de Physique, Universit\'{e} de Sherbrooke, Sherbrooke J1K2R1 Qu\'{e}bec, Canada}

\author{Elie Genois}
\affiliation{Institut Quantique and D\'{e}partement de Physique, Universit\'{e} de Sherbrooke, Sherbrooke J1K2R1 Qu\'{e}bec, Canada}

\author{Catherine Leroux}
\affiliation{Institut Quantique and D\'{e}partement de Physique, Universit\'{e} de Sherbrooke, Sherbrooke J1K2R1 Qu\'{e}bec, Canada}

\author{Christoph Hellings}
\affiliation{Department of Physics, ETH Zurich, CH-8093 Zurich, Switzerland}

\author{Stefania Lazar}
\affiliation{Department of Physics, ETH Zurich, CH-8093 Zurich, Switzerland}

\author{Francois Swiadek}
\affiliation{Department of Physics, ETH Zurich, CH-8093 Zurich, Switzerland}

\author{Johannes Herrmann}
\affiliation{Department of Physics, ETH Zurich, CH-8093 Zurich, Switzerland}

\author{Graham J. Norris}
\affiliation{Department of Physics, ETH Zurich, CH-8093 Zurich, Switzerland}

\author{Christian Kraglund Andersen}
\thanks{Current affiliation: QuTech and Kavli Institute for Nanoscience, Delft University of Technology, Delft 2628 CJ, Netherlands}
\affiliation{Department of Physics, ETH Zurich, CH-8093 Zurich, Switzerland}

\author{Markus Müller}
\affiliation{Institute for Quantum Information, RWTH Aachen University, Aachen D-52056, Germany}
\affiliation{Peter Grünberg Institute, Theoretical Nanoelectronics, Forschungszentrum Jülich, Jülich D-52425, Germany}

\author{Alexandre Blais}
\affiliation{Institut Quantique and D\'{e}partement de Physique, Universit\'{e} de Sherbrooke, Sherbrooke J1K2R1 Qu\'{e}bec, Canada}
\affiliation{Canadian Institute for Advanced Research, Toronto, ON, Canada}

\author{Christopher Eichler}
\affiliation{Department of Physics, ETH Zurich, CH-8093 Zurich, Switzerland}

\author{Andreas Wallraff}
\affiliation{Department of Physics, ETH Zurich, CH-8093 Zurich, Switzerland}
\affiliation{Quantum Center, ETH Zurich, 8093 Zurich, Switzerland}

\date{\today}

\begin{abstract}
  Quantum computers hold the promise of solving computational problems which are intractable using conventional methods \cite{Preskill2018}.
  For fault-tolerant operation quantum computers must correct errors occurring due to unavoidable decoherence and limited control accuracy~\cite{Shor1996}.
  Here, we demonstrate quantum error correction using the surface code, which is known for its exceptionally high tolerance to errors \cite{Kitaev2003, Dennis2002, Raussendorf2007, Bombin2009}.
  Using 17 physical qubits in a superconducting circuit we encode quantum information in a distance-three logical qubit building up on recent distance-two error detection experiments \cite{Andersen2020b, Marques2021, Chen2021p}.
  In an error correction cycle taking only $1.1\,$µs, we demonstrate the preservation of four cardinal states of the logical qubit.
  Repeatedly executing the cycle, we measure and decode both bit- and phase-flip error syndromes using a minimum-weight perfect-matching algorithm in an error-model-free approach and apply corrections in postprocessing.
  We find a low error probability of $3\,\%$ per cycle when rejecting experimental runs in which leakage is detected.
  The measured characteristics of our device agree well with a numerical model.
  Our demonstration of repeated, fast and high-performance quantum error correction cycles, together with recent advances in ion traps \cite{Ryan-Anderson2021}, support our understanding that fault-tolerant quantum computation will be practically realizable.
\end{abstract}

\maketitle


The surface code \cite{Bravyi1998, Dennis2002} is a planar realization of Kitaev's toric code \cite{Kitaev2003} which uses topological features of a qubit lattice to correct errors in quantum information processing systems.
This code is a prominent contender to reach fault-tolerant quantum computation because of its high error threshold of about $1\,\%$ against quantum circuit noise \cite{Wang2003a, Raussendorf2007} and its compatibility with 2D architectures.
The surface code belongs to the family of stabilizer codes \cite{Gottesman1997, Terhal2015n} which encode quantum information into a joint subspace of definite parities on a set of physical data qubits to form a logical qubit. Errors are detected using measurements of auxiliary qubits to extract parity information without collapsing the logical qubit state.
The fault-tolerant operation of a quantum computer requires repeated detection and correction of both bit- and phase-flip errors on data qubits. With an increasing number of physical qubits and thus an increasing code distance $d$ the number of errors $\lfloor (d-1)/2 \rfloor$ which can at least be detected and corrected per error-correction cycle increases, making the code more resilient when error rates are sufficiently low.

Error correction limited to a single type of error has been realized with repetition codes in nuclear magnetic resonance \cite{Moussa2011}, trapped ions \cite{Schindler2011}, nitrogen-vacancy centers \cite{Waldherr2014} and superconducting circuits \cite{Kelly2015, Chen2021p}. In single-cycle experiments, fault-tolerant stabilizer measurements and correction of both types of errors have been demonstrated with the five-qubit code and the Bacon-Shor code \cite{Knill2001a, Abobeih2021, Egan2021,Hilder2021}. Recently, error detection in a distance-two surface code has been realized with seven qubits \cite{Andersen2020b, Marques2021, Chen2021p}, and only very recently, repeated stabilizer-based error correction has been demonstrated with a distance-three color code in a trapped ion system \cite{Ryan-Anderson2021}.

Correction of both bit- and phase-flip errors requires at least a distance-three code. In combination with fault-tolerant circuits for error syndrome measurements, this guarantees that any single error on any of the constituent data and auxiliary qubits or operations can be corrected \cite{Horsman2012,Terhal2015n}.
While the work we discuss here focuses on digital encoding of quantum information, continuous variable encoding, for example in harmonic oscillator states, constitutes an alternative approach to quantum error correction~(QEC), see for example Refs.~\onlinecite{Ofek2016, Hu2019a, Fluehmann2019, CampagneIbarcq2020}.

\section*{A Distance-Three Surface Code in Superconducting Circuits}

\begin{figure*}[t!] 
  \centering
  \includegraphics{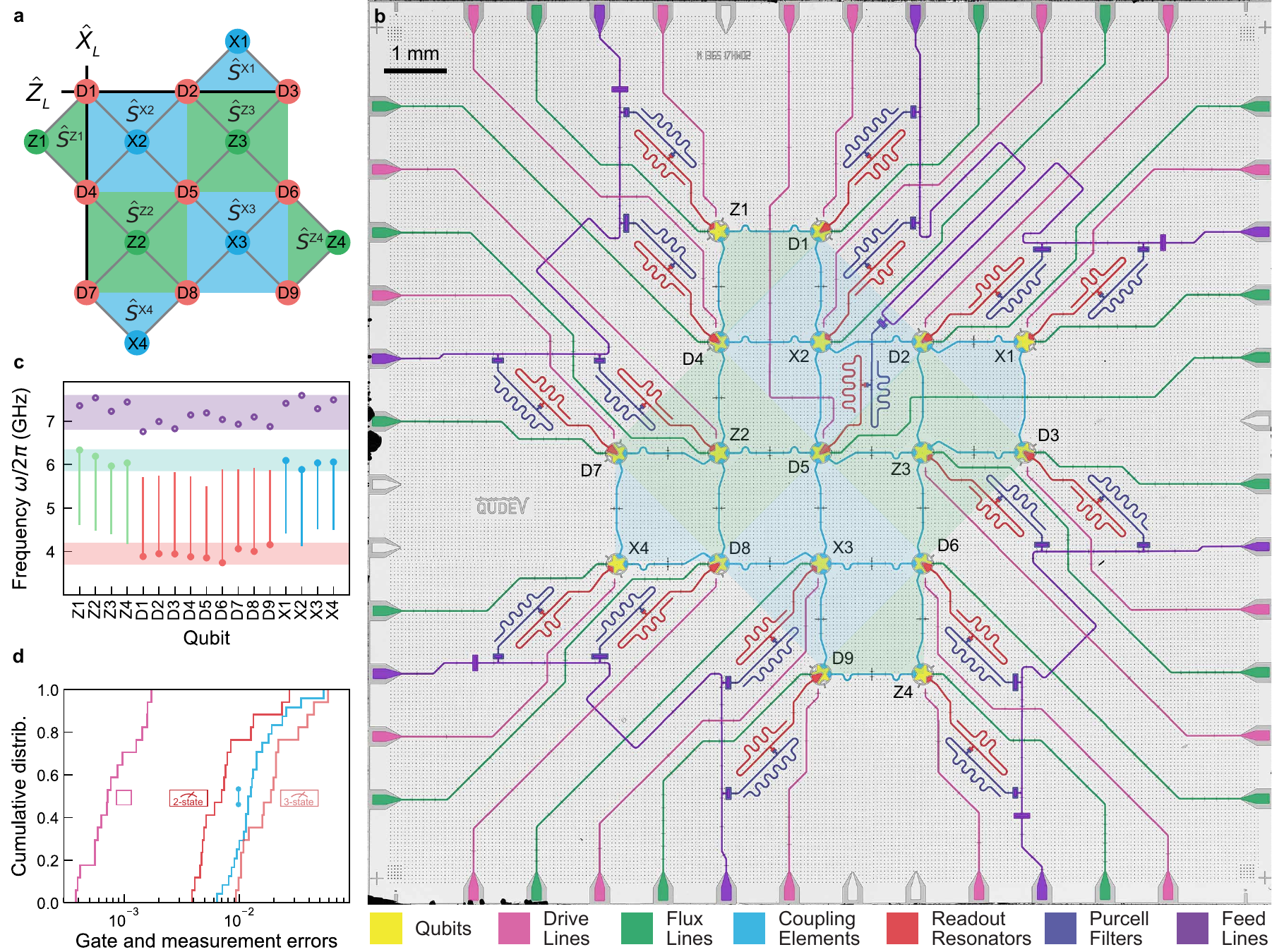}
  \caption{Device concept, architecture and performance.
  \textbf{a}~Conceptual representation of the distance-three surface code consisting of data qubits (red circles), Z-type (green circles) and X-type auxiliary qubits (blue circles), with their connectivity indicated by gray lines. The data qubits participating in the weight-three logical operators $\lopz$ and $\lopx$ are indicated by solid black lines. Green (blue) plaquettes indicate X-type (Z-type) stabilizer circuits.
  \textbf{b}~False-color micrograph of the device realizing the concept in \textbf{a} with 17 transmon qubits, see legend for circuit elements and text for details. The qubit lattice is rotated by 45 degrees with respect to  \textbf{a}. \textbf{c}~Frequency arrangement in three distinct bands for idling data qubits (red circles),  idling Z/X-type auxiliary qubits (green/blue circles), and readout resonators (violet open circles). The qubit-frequency tuning ranges are indicated by vertical bars.
  \textbf{d}~Cumulative distributions (integrated histograms) of single-qubit gate (pink), simultaneous two-qubit gate (cyan), two-state (red) and three-state readout errors (light red).}
  \label{fig:device}
\end{figure*}

Experimentally realizing a distance-three surface code requires nine data qubits and eight auxiliary qubits, also referred to in the literature as ancilla or measurement qubits \cite{Bombin2007, Horsman2012, Tomita2014a}. The qubits are arranged in a diagonal, planar square lattice, the edges of which are shown in gray in the schematic Fig.~\ref{fig:device}a. The data qubits D$j$,~$j=1 \ldots 9$, (red dots) form a $3\times 3$ array and are interlaced with auxiliary qubits $Ai$, labeled X$i$ (blue) and Z$i$,~$i=1 \ldots 4$ (green). We realized this  arrangement in a superconducting circuit using 17 transmon qubits \cite{Koch2007} (yellow) capacitively coupled to each other along the edges of the square array with $\sim 1$~mm long coplanar waveguide segments (turquoise), see Fig.~\ref{fig:device}b. We discuss the fabrication of this device in \cref{app:device_fabrication}.

Using the auxiliary qubits X$i$ and Z$i$ we measure the parity of the neighboring two or four data qubits D$j$, which are located at the vertices of the blue and green plaquettes, in the X or Z basis (Fig.~\ref{fig:device}a,b).
\natmethods{In a Z-basis measurement, if an odd number of the data qubits involved in the parity operator under consideration is in the $|1\rangle$-state the auxiliary qubit state is flipped. On the other hand, if an even number of data qubits is in  $|1\rangle$ the auxiliary qubit state remains unchanged. The equivalent is true in the X basis for an even or odd number of data qubits in the $|-\rangle$-state. Here, $|0\rangle$, $|1\rangle$ are the transmon qubit ground and first excited states, and $|\pm\rangle=\left(|0\rangle \pm |1\rangle\right)/\sqrt{2}$ are their superpositions.
To map the parity of the data qubits D$j$ onto the corresponding auxiliary qubit, we effectively use a sequence of controlled-not gates with the data qubits as control and the auxiliary qubit as target, and subsequently measure the state of the auxiliary qubit in the Z basis using single-shot readout. In the Z basis, a single bit-flip error of any individual data qubit leads to a change of parity, as does a single phase-flip in the X basis. Hence, measurements of changes of data-qubit parities allow us to detect and identify phase-flip or bit-flip errors as long as they occur sufficiently rarely \cite{Fowler2012}.}
These parity measurements are also referred to as stabilizer measurements \cite{Gottesman1997, Terhal2015n}. The corresponding mutually-commuting weight-two (or weight-four) stabilizer operators $\sx{i} = \prod_{j=1}^{2(4)}\px_j$ and $\sz{i} = \prod_{j=1}^{2(4)}\pz_j$ of the surface code are products of two (or four) Pauli-$\px$ or -$\pz$ operators of the data qubits $j$ located at the vertices of a given data-qubit plaquette. Measurement outcomes $s^{Ai} = \pm 1$ of individual stabilizers $\sai$ are extracted from the observed change of auxiliary qubit state from one cycle to the next and indicate even or odd parity, respectively.

\begin{figure}[t] 
  \centering
  \includegraphics{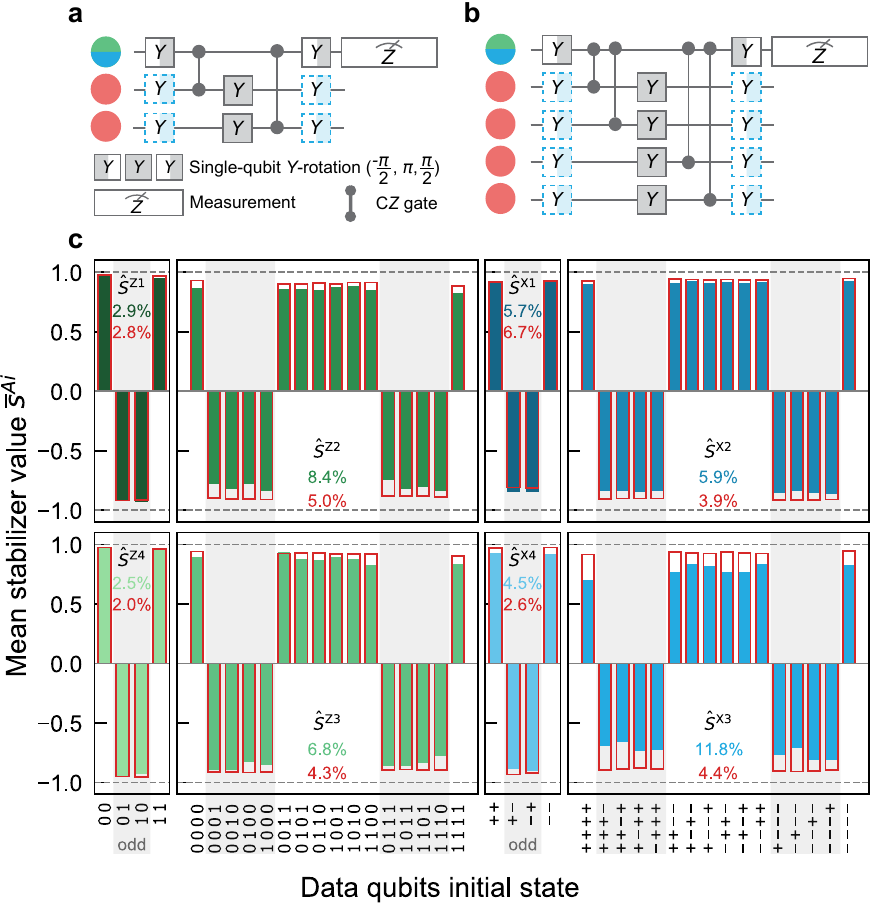}
  \caption{Stabilizer circuits and their characterization.
  Quantum circuit diagrams for \textbf{a}~weight-two and \textbf{b}~weight-four stabilizers acting between data qubits located on the vertices of a plaquette (red circles) and the corresponding auxiliary qubit used for Z-type (green) or X-type (blue circle) stabilizer measurements, where the latter require a basis change (dashed-blue squares), see text for details.
  \textbf{c}~Measured (filled bars), and simulated (red wireframes) average stabilizer values $\overline{s}^{Ai}$ versus data qubit input state, ordered by number of excitations. Percentages are the experimental (blue, green) and simulated (red) error of $\overline{s}^{Ai}$. Gray background indicates odd and white even parity.
  }
  \label{fig:stabilizer}
\end{figure}

In our experiments, the stabilizer gate sequence is realized as two or four controlled-phase (CZ) gates \cite{Strauch2003, DiCarlo2010, Negirneac2021} (see Methods \cref{methods_sec:cz_gates}) between data and auxiliary qubits, operated in a high and a low frequency band (see Fig.~\ref{fig:device}c), respectively, combined with initial and final $\pi/2$-rotations on the auxiliary qubits (Fig.~\ref{fig:stabilizer}a,b). The gate sequence for measuring $\sx{i}$ contains additional initial and final $\pi/2$-rotations acting on the data qubits, implementing a basis change from the Z to the X basis (blue dashed squares in Fig.~\ref{fig:stabilizer}a and b). We apply echo pulses to the data qubits in the middle of the gate sequence to reduce dephasing of the data qubits and residual coherent coupling to spectator qubits~\cite{Krinner2020}.


The 24 pairwise \CZs{} have a mean duration of $98(7)~$ns, including two conservatively chosen 15-ns-long buffers at the beginning and the end, and display a mean gate error of 0.015(10). The gate error histogram, displayed as an integrated (cumulative) distribution,
shows variations of about a factor of four in two-qubit gate error (Fig~\ref{fig:device}d). \natmethods{We determine the two-qubit gate error from interleaved randomized benchmarking experiments with sets of three gates executed in parallel, as employed in our realization of the surface code cycle. Time-varying microscopic defects in our device have a detrimental influence on two-qubit gate performance and are responsible for outliers in the gate error distribution (\cref{app:two_qubit_gates}).}
Single-qubit gates displaying a mean error of 0.0009(4) are realized by applying short resonant microwave pulses to each qubit individually through a dedicated drive line (pink coplanar waveguide in Fig.~\ref{fig:device}b). We determined the single-qubit gate fidelities in randomized benchmarking experiments. We discuss the experimental setup used to realize these gates in \cref{app:experimental_setup}.

A key element of individual stabilizer measurements are fast and high-fidelity measurements of auxiliary qubit states while leaving data qubit states unaffected \cite{Negnevitsky2018, Andersen2019, Bultink2020} (see Methods \cref{methods_sec:readout}).
Accurate stabilizer measurements using mid-cycle qubit readout on timescales comparable to or shorter than the cumulated gate times per error correction cycle also contribute to maximizing the performance of error detection and correction in our surface code implementation as a whole.

With our readout scheme we discriminate the two computational qubit states and a leakage state, which if undetected or uncorrected for is detrimental to any surface code implementation \cite{Alferis2007, Fowler2013, Ghosh2015a, Suchara2015, Bultink2020, Varbanov2020}. We achieve a mean readout assignment error of $0.009(7)$ when discriminating the computational states only (two-state readout) and of $0.022(14)$ when discriminating the computational states and the leakage state (three-state readout), see \cref{app:readout_characterization}. The corresponding cumulative distribution for two-state readout exhibits performance variations on the device of about a factor of two when disregarding two outliers, while the distribution for three-state readout shows variations larger by about a factor of two (Fig~\ref{fig:device}d).

With all elements in place for realizing a surface code, we first characterize the measurements of individual $\sai$ stabilizers. To do so we prepare the data qubits D$j$ of a given weight-two or weight-four plaquette sequentially in each one of its $2^2=4$ or $2^4=16$ basis states composed of $|0\rangle$ and $|1\rangle$ for $\sz{i}$ and $|+\rangle$ and $|-\rangle$ for $\sx{i}$.
In the beginning of each experiment all qubits are initialized by heralding the ground state $|0\rangle$ from single-shot readout.

For each input state, we compute the mean values $\overline{s}^{Ai}$ from $\sim 4\times 10^4$ measurements of $s^{Ai}$ (colored bars in Fig.~\ref{fig:stabilizer}c) and find good qualitative agreement with master equation simulations (red outlines), see Appendices~\ref{app:numerical_simulations} and \ref{app:stabilizers} for details. Here, $+1/-1$ indicate even/odd parity of the measured state. The colored percentage values display the corresponding experimental and simulated errors of the stabilizer measurements. We attribute the differences between measurements and simulation mostly to two-qubit gate errors due to microscopic defect modes changing their frequency on timescales of hours or days (Appendix~\ref{app:two_qubit_gates}).

Having verified that all stabilizer measurements perform at high quality levels individually, we combine the stabilizer measurements into a surface code cycle. Executing this cycle once, we prepare one of the four cardinal logical qubit states. Executing the cycle multiple times, we stabilize the logical states and investigate the performance of our realization of the code.

\section*{The Surface Code Cycle and State Initialization}

At the beginning of each experimental sequence, we
prepare the nine data qubits in either one of the product states $\ket{0}^{\otimes 9}$ and $\lopx{}\ket{0}^{\otimes 9}$ ($\ket{+}^{\otimes 9}$ and $\lopz{}\ket{+}^{\otimes 9}$)
to begin the process of initializing the cardinal logical qubit states $\lz$ and $\lo$ ($\lp$ and $\lm$). The cardinal states are eigenstates of the eight stabilizer operators $\sai$ and $\pm 1$ eigenstates of the logical Pauli operators which we choose as $\lopz = \pz_1\pz_2\pz_3$ and $\lopx = \px_1\px_4\px_7$, see solid black lines in Fig.~\ref{fig:device}a. As required, $\lopz$ and $\lopx$ commute with all stabilizers and anti-commute with each other.
Since each of the prepared product states is an equal superposition of $16$ equivalent instances of the target logical state, executing a single quantum error correction cycle once deterministically initializes the target logical state in the stabilizer eigenspace corresponding to the measurement outcome of the stabilizers (\cref{app:logical_state_fidelity}).

\begin{figure}[t] 
\centering
\includegraphics[width=\columnwidth]{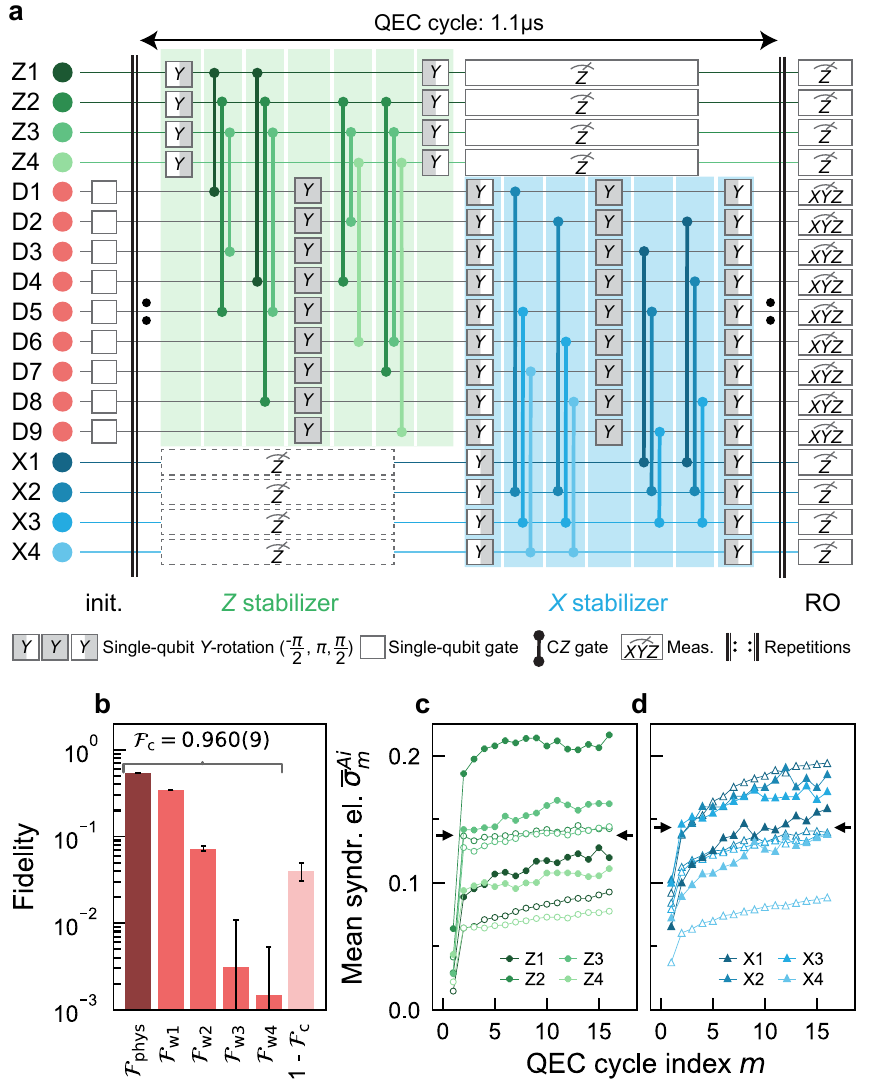}
  \caption{The surface code cycle, fidelity of logical state initialization, and average error syndromes.
  \textbf{a}~Quantum circuit used to initialize and (repeatedly) error correct the distance-3 surface code logical qubit. Green (blue) shaded circuit elements represent the parallel execution of the four Z-type (X-type) stabilizer circuits. Empty squares indicate single-qubit rotations on data qubits. In the first cycle the X$i$ auxiliary qubits are not measured (dashed boxes).
  \textbf{b}~Fidelity of the prepared logical state \lz{} (dark red bar), all correctable states differing from \lz{} by one to four Pauli errors (red bars) and all uncorrectable states (light red bar).
  Average syndrome elements $\overline{\sigma}_m^{Ai}$ as a function of the cycle index $m$ for \textbf{c} the \lopz{}  (filled circles) and \textbf{d} the \lopx{} (filled triangles) preservation experiment. Open symbols are simulations. The horizontal black arrows indicate the average syndrome element over all stabilizers and cycles, see main text for details.}
\label{fig:stabilizers_state_init}
\end{figure}

In a single surface code cycle, we first execute all gate operations implementing the four $\sz{i}$ stabilizer measurements. We realize the necessary two-qubit gates in four time steps in each of which we execute three CZ gates simultaneously.  Parallelizing stabilizer execution is a key technical requisite for scalable quantum error correction, in particular for operation of larger-distance codes. \natmethods{ The two-qubit gates are accompanied by a set of single-qubit gates applied to all auxiliary qubits in a leading and a trailing time step, and a dynamical decoupling pulse applied to all data qubits at a central time step (Fig.~\ref{fig:stabilizers_state_init}a). We choose the order of gate operations to provide resilience against single auxiliary qubit errors and to avoid interactions with microscopic defect modes (\cref{app:gate_sequence}).}
The gate execution is followed by readout of the Z-type auxiliary qubits to complete the Z-type stabilizer measurement, see Fig.~\ref{fig:stabilizers_state_init}a. Simultaneous with the Z-type auxiliary qubit readout,  we start executing the X-type stabilizer circuits, which are equivalent up to an additional basis change of the data qubits. This allows us to execute the $\sz{i}$ and $\sx{i}$ stabilizer measurements in a parallel, pipelined approach \cite{Versluis2017}, see Fig.~\ref{fig:stabilizers_state_init}a for a full circuit diagram and \cref{app:pulse_sequence} for a full pulse sequence.

Thanks to the pipelined approach combined with short qubit-readout times, we are able to execute a single quantum error correction cycle in a cycle time as short as $t_c = 1.1\,$µs, setting the rate at which we detect errors for their subsequent correction. A short cycle time is essential for any error correction code as it determines the performance of the code at a given physical qubit coherence time. Moreover, a short cycle time reduces the execution time of error-corrected quantum algorithms \cite{Burg2021,Babbush2021}.

To maximize performance, we have designed our device with parameters minimizing leakage on data qubits. We also detect the residual leakage on auxiliary qubits during the execution of each cycle, and on data qubits after the last cycle, using our three-state readout (see Methods \cref{methods_sec:leakage}).
In our experiments we reject all instances of detected leakage events. The degree to which leakage rejection affects the retained fraction of experimental runs is discussed in \cref{app:leakage_rejection}.

To characterize our logical-state initialization scheme, we determine the fidelity of the prepared state with respect to the target logical state by measuring the $2^9$ Pauli strings which form the basis of the 9-data-qubit logical state~\cite{Nigg2014b,Abobeih2021}. For \lz{}
we find a quantum state fidelity $\mathcal{F}_\mathrm{phys} = \trace(\rho |0\rangle_{\mathrm{L}}\langle 0|_{\mathrm{L}}) =  54.0(1)\%$ (dark red bar in \cref{fig:stabilizers_state_init}b) when using our leakage-detection scheme and when correcting for readout errors on data qubits.
Considering only errors in the logical subspace~\cite{Andersen2020b}, we find a logical fidelity  $\mathcal{F}_\mathrm{L} =  \mathcal{F}_\mathrm{phys} / P_\mathrm{L} = 99.6(2)\,\%$ where $P_\mathrm{L} = 54.2(1)\,\%$ is the experimentally measured probability of preparing a state in the logical subspace (\cref{app:logical_state_fidelity}).
Both $\mathcal{F}_\mathrm{phys}$ and $P_\mathrm{L}$ are smaller than in a distance-two surface code, see Ref.~\onlinecite{Andersen2020b} for example, since the two quantities are expected to decrease with increasing distance $d$ at constant physical error rate.

To further evaluate the performance of our logical state initialization, we analyze the fidelity of the prepared state with respect to subspaces of states which our surface code implementation can in principle correct.
The errors which are correctable by the distance-three surface code include all single-qubit Pauli (weight-one) errors $\hat X_j$, $\hat Y_j$, or  $\hat Z_j$ on any data qubit $j$ and a subset of higher-weight errors. We compute the fidelity of the experimentally prepared state with respect to correctable subspaces including states which are equivalent to the target state up to Pauli errors of weight $i$ up to $i=4$ and find $(\mathcal{F}_\mathrm{w1},\mathcal{F}_\mathrm{w2},\mathcal{F}_\mathrm{w3},\mathcal{F}_\mathrm{w4}) = [34.3(0),\,7.3(5), \,0.3(8), \,0.1(4)]\,\%$, see \cref{app:logical_state_fidelity} and Ref.~\cite{Abobeih2021}.
Hence, the prepared initial state has a fidelity of $\mathcal{F}_\textrm{c} =\mathcal{F}_\textrm{phys} + \sum_{i=1}^{4}\mathcal{F}_{\textrm{w}i} = 96.0(9)\,\%$ with respect to states which are in principle correctable in our realization of the surface code. We also note that weight-one errors account for the majority of errors on data qubits in the \lz{} state initialization, with higher weight errors having a largely reduced probability of occurrence (\cref{fig:stabilizers_state_init}b).

\section*{Repeated Quantum Error Correction}\label{sec:QEC}

Once the first quantum error correction cycle completes the logical state initialization, we make use of all subsequent cycles for logical state preservation. In our experiments, we preserve the cardinal logical qubit states for up to $n=16$ cycles. In each cycle $m=1,..,n$ we extract eight stabilizer values $s_m^{Ai}$. Changes in stabilizer values signal the occurrence of errors and are used to construct error syndromes $\sigma_m$ consisting of eight syndrome elements $\sigma_m^{Ai}=(1-s_m^{Ai} \times s_{m-1}^{Ai})/2$.  The elements $\sigma_m^{Ai}$ are inferred in each cycle from the current $(m)$ and the previous ($m-1$) measured stabilizer values with $\sigma_m^{Ai}=1(0)$ indicating an error (no error), respectively \cite{Kelly2015}.

We collectively process successive syndromes $\sigma_m$ to determine which data and auxiliary qubits have most likely suffered an error \cite{Dennis2002, Fowler2012, Terhal2015n}.
\natmethods{Specifically, we construct a graph in which the syndrome elements are displayed at the auxiliary qubit locations along two spatial coordinates for each cycle index $m$ which forms the temporal coordinate. Spatial and temporal correlations between non-zero syndrome elements correspond to data and auxiliary qubit errors, respectively. If the overall error rate is sufficiently low, we obtain a low density of non-zero syndrome elements, or equivalently, mean syndrome element values $\overline{\sigma}_m^{Ai}\ll 1$, with the mean taken over experimental realizations. In that case, the underlying errors can be decoded with low ambiguity (\cref{app:decoding}).}

Executing $n$ consecutive error correction cycles and rejecting runs in which leakage has been detected, we record stabilizer measurement outcomes and construct syndromes from their values, the averages of which are shown in Fig.~\ref{fig:stabilizers_state_init}c,d.
When averaging the syndromes over all individual elements and time, we find that the average syndrome element $\overline{\sigma} = 0.14 \ll 1$ is small (see arrows in Fig.~\ref{fig:stabilizers_state_init}c,d), indicating that errors are rare and, therefore, allowing for efficient error detection and correction~\cite{Chen2021p}.
\natmethods{All syndrome elements $\overline{\sigma}_m^{Ai}$ averaged over repetitions of the experiments are approximately constant as a function of $m$ for $m \geq 2$, see Fig.~\ref{fig:stabilizers_state_init}c,d obtained for preserving \lz{},\lo{} and \lp, \lm, respectively. We attribute the small remaining increase of $\overline{\sigma}_m^{Ai}$ with $m$ to the fact that, in the absence of auxiliary qubit reset, auxiliary qubits initially prepared in the ground state tend to an asymptotic probability of $0.5$ to be in the excited state after $m$ cycles. As a result, with increasing $m$, auxiliary qubits suffer from larger decoherence during readout and during the subsequent idling periods of about $150\,$ns before the start of the next quantum-error-correction cycle.
Our numerical simulations show the same feature (open symbols in Fig.~\ref{fig:stabilizers_state_init}c,d).
For $m=1$, the averaged syndrome elements $\overline{\sigma}_1^{Ai}$ are reduced because the corresponding reference stabilizer values are computed from the initial data qubit product state, which we prepare with high fidelity. In the first cycle, the four values of $\overline{\sigma}_1^{\textrm{Z}i}$ are smaller than $\overline{\sigma}_1^{\textrm{X}i}$ because a quantum-error-correction cycle starts with measurements of $\sz{i}$ and errors thus accumulate only during half a cycle.}

To determine the performance of our distance-three surface code, we extract the logical error per cycle when preserving eigenstates of the logical qubit operators $\lopz$ and $\lopx$ versus the number of executed cycles $n$. For each sequence of cycles, we decode the error syndromes, including a syndrome determined from the final data qubit readout. After the $n^{\rm{th}}$ cycle we perform a projective readout of the final data qubit state in the Z or X basis from which we determine the eigenvalue $z_{\rm L}=\pm 1$ of $\lopz$ or $x_{\rm L}=\pm 1$ of $\lopx$.

\begin{figure}[!t] 
  \centering
  \includegraphics{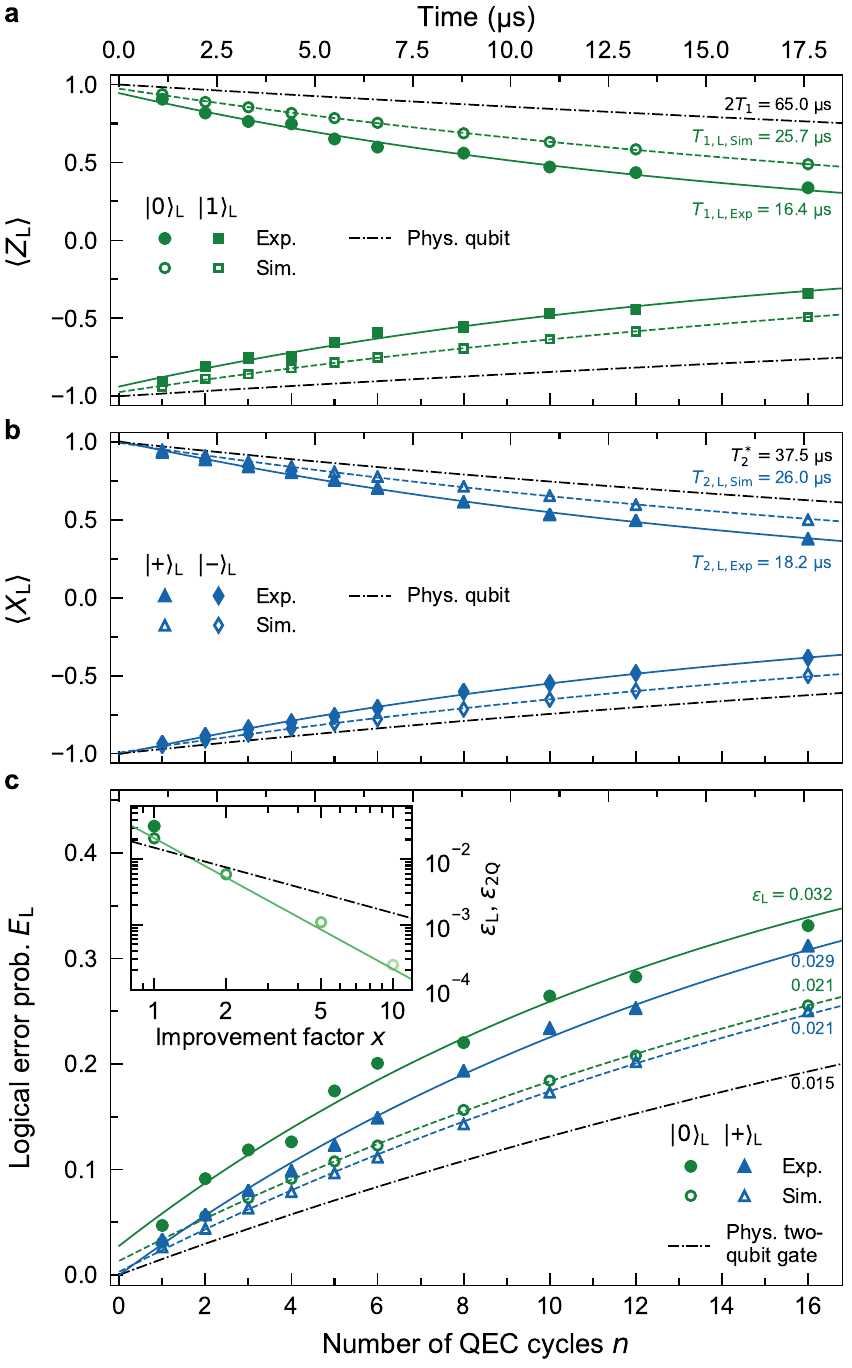}
  \caption{Logical state preservation and error per cycle.
  \textbf{a}~Experimentally determined (full symbols) and simulated (open symbols) expectation value of the \lopz{} operator for prepared $\lz$ (circles) and $\lo$ (squares) and exponential fit (solid and dashed line). For reference, twice the average physical qubit decay is shown (dash-dotted line).  The extracted decay times are indicated on the right.  \textbf{b}~Corresponding data sets for \lopx{}, $\lp$ (triangles) and $\lm$ (diamonds). For reference, the average physical qubit coherence decay is indicated (dash-dotted line).
  \textbf{c}~Logical error probability $E_{\rm L}$ as a function of the number $n$ of error correction cycles for $\lz$ and $\lp$ and the extracted error per cycle $\epsilon_L$ indicated on the right. Same symbols as in \textbf{a} and \textbf{b}. The dash-dotted line shows the physical two-qubit gate error accumulated over $n$ cycles, for reference. The inset shows simulation results for $\epsilon_{\rm L}$ in state $\lz$ (open, green circles) for an assumed homogeneous distribution of decoherence rates, gate and readout errors, reduced by factors of $x = 2, 5$ and $10$, see text for details. The green solid line is a fit to $1/x^2$.
  }
  \label{fig:state_preservation}
\end{figure}

Decoding the error syndromes and applying potential corrections to $z_{\rm L}$ or $x_{\rm L}$, we compute the mean logical qubit expectation values $\overline{z}_{\rm L}=\langle \lopz \rangle$ and $\overline{x}_{\rm L}=\langle \lopx \rangle$ as a function of $n$ from a total of $10^6$ experimental runs, where the available data is reduced by ground state heralding before and leakage rejection during each run.
\natmethods{We decode the error syndromes using a minimum-weight perfect matching algorithm \cite{OBrien2017, Edmonds1965}. We determine the weights in an error-model-free approach by inferring the errors per cycle from the measured data using a correlation analysis of the syndromes as described in \cref{app:decoding} and Refs.~\cite{Spitz2018, Chen2021p}.
The correction of an error, initiated by analysing all cycles in postprocessing, takes the form of changing the sign of the logical qubit operator values $z_{\rm L}$ and $x_{\rm L}$ when indicated by the decoder. We note that for correcting $\lopz$ it is sufficient to decode only syndromes $\{\sigma_m^{\textrm{Z}i}\}$, or, equivalently, only $\{\sigma_m^{\textrm{X}i}\}$ for $\lopx$.}

We observe an exponential decay of $\langle \lopz \rangle$ and $\langle \lopx \rangle$ with $n$ (solid symbols in Fig.~\ref{fig:state_preservation}a,b).
From the logical qubit operator expectation values we extract the logical error probability $E_{\mathrm{L}}=\left(1-|\langle Z_{\mathrm{L}}\rangle|\right)/2$ (solid symbols in Fig.~\ref{fig:state_preservation}c) as a function of $n$ and find a small per cycle error probability of $\epsilon_{\mathrm{L}} = \left[1-{\rm exp}\left(-t_{\rm c}/T_{1,\mathrm{L}}\right)\right]/2 \approx t_{\rm c}/2T_{1,\mathrm{L}} = 0.032(1)$ also indicated on the right hand side of the corresponding data set in Fig.~\ref{fig:state_preservation}c. Equivalently, we obtain $\epsilon_{\mathrm{L}} = \left[1-{\rm exp}\left(-t_{\rm c}/T_{2,\mathrm{L}}\right)\right]/2 \approx t_{\rm c}/2T_{2,\mathrm{L}} = 0.029(1) $ for $\langle X_{\mathrm{L}}\rangle$. We find that both the coherence time of $T_{2,\mathrm{L}}=18.2(5)\,$µs and the lifetime $T_{1,\mathrm{L}}=16.4(8)\,$µs of the logical qubit, as extracted from the decay curves of $\langle \lopz \rangle$ and $\langle \lopx \rangle$, are much longer than the duration of the quantum error correction cycle $t_c=1.1\,$µs in our implementation of the surface code. \natmethods{In these experiments, we reject detected leakage events which occur with small probabilities on auxiliary and data qubits. As detailed in Appendix~\ref{app:leakage_rejection}, the retained fraction of data per cycle amounts to 92~\%.}

We note that the logical coherence time is only about a factor of two lower than the mean physical coherence time $\overline{T_2^*}=37.5\,$µs of all 17 qubits while the logical relaxation time is about a factor four lower than twice the mean physical energy relaxation time $2 \overline{T_1}=65.0\,$µs, as indicated in Fig.~\ref{fig:state_preservation}a,b. In the latter case the factor of two assumes an infinite lifetime of the physical qubit ground state for the comparison. We also note that, within errors bars, the lifetime of the states $\lz$ and $\lo$, and the coherence time of $\lp$ and $\lm$ are identical. This is expected since all cardinal states of the logical qubit have the same number of qubits in the excited state and our dynamical decoupling scheme alternates between the basis states which are therefore similarly affected by decoherence.

\section*{Performance Assessment and Projection}

We compare the state preservation experiments with numerical simulations using a Monte Carlo wavefunction method (open symbols in Fig.~\ref{fig:state_preservation}a,b,c). The model underlying the simulations (\cref{app:numerical_simulations}) uses the measured
coherence times, interaction rates and readout errors of the device as inputs.
We find that despite the complexity of the quantum error correction cycle the measured expectation values $\langle \lopx \rangle$ and $\langle \lopz \rangle$, when rejecting leakage, come close to the simulated values (open symbols in Fig.~\ref{fig:state_preservation}a,b).
The logical lifetimes extracted from exponential fits (dashed lines in Fig.~\ref{fig:state_preservation}a,b) to the simulated expectation values are approximately $26\,$µs providing an upper bound for the performance achievable with the specified device parameters.

Since the numerical simulations model the performance of our quantum device well, we use the model to project how future improvements in gate and readout fidelities are expected to reduce the logical error per cycle $\epsilon_L$. To free the projection from device-specific spread in qubit parameters, we use the average over all 17 qubits as uniform parameters (see Tab.~\ref{tab:qubit_measured_parameters}), and find good agreement with the results obtained from the qubit-specific model.
We then uniformly reduce all physical error parameters of the numerical model by a factor $x$, repeat the simulations of $\langle \lopx \rangle$ and $\langle \lopz \rangle$ \emph{vs}~$n$, and extract the mean logical error per cycle $\epsilon_L$ as a function of the improvement factor $x$. We find that the simulated error per cycle scales to a good approximation as $1/x^2$, see inset in Fig.~\ref{fig:state_preservation}c, as expected for a code of distance three \cite{Fowler2012}. For reference, we plot the scaled two-qubit physical error per cycle $\epsilon_{\rm{2Q}}$ in the same plot (dash-dotted line).

A metric commonly used to assess the performance of quantum error correction compares the logical error per cycle $\epsilon_{\rm L}$ to the dominant error on the physical level, typically the two-qubit gate error $\epsilon_{\rm{2Q}}$ \cite{Ryan-Anderson2021,Trout2018}.
Such a comparison is particularly relevant in architectures in which the logical two-qubit gate error is dominated by errors $\epsilon_L$ in the quantum error correction cycles belonging to or following the logical two-qubit gate operation. The number of required cycles scales in general with $d$ in planar architectures and in architectures allowing for transversal execution of logical two-qubit gates \cite{Raussendorf2007, Bombin2009, Horsman2012, Landahl2014, Gutierrez2019}.
The good agreement between measured ($\sim 3\,\%$) and simulated ($\sim 2\,\%$) logical errors $\epsilon_{\rm L}$ together with their simulated quadratic scaling suggests that the break-even of per-cycle logical errors with two-qubit gate errors may be in reach for modest improvements of device performance, when employing leakage detection or correction.

In our experiments, we demonstrate the viability of realizing quantum error correction in the surface code by detecting errors during the error correction cycle, and decoding the error syndromes and correcting for errors in postprocessing, which is sufficient in a quantum memory setting. Next generation experiments will provide the capability of correcting errors during the cycle using real-time decoding~\cite{Ryan-Anderson2021} and fast in-sequence feedback~\cite{Andersen2019}, implemented with dedicated digital electronics. Feedback will also enable the mid-cycle suppression of leakage~\cite{McEwen2021a}, for example by auxiliary qubit reset.
Realizing larger surface code lattices while improving the performance of their components and demonstrating exponential suppression of logical errors with increasing code distance are upcoming important steps toward achieving the long-term goal of fault-tolerant quantum computation.

\section*{Acknowledgments}
The authors are grateful for valuable discussions with Quentin Ficheux and Cristóbal Lledó. The authors acknowledge the contributions of Richard Boell to the experimental setup and of Michael Kerschbaum for early work on the two-qubit gate implementation.

The team in Zurich acknowledges financial support by the Office of the Director of National Intelligence (ODNI), Intelligence Advanced Research Projects Activity (IARPA), via the U.S. Army Research Office grant W911NF-16-1-0071, by the EU Flagship on Quantum Technology H2020-FETFLAG-2018-03 project 820363 OpenSuperQ, by the National Centre of Competence in Research Quantum Science and Technology (NCCR QSIT), a research instrument of the Swiss National Science Foundation (SNSF), by the SNFS R'equip grant 206021-170731 and by ETH Zurich. S.K. acknowledges financial support from Fondation Jean-Jacques \& Felicia Lopez-Loreta and the ETH Zurich Foundation. The work in Sherbrooke was undertaken thanks in part to funding from NSERC, Canada First Research Excellence Fund and ARO W911NF-18-1-0411, the Ministère de l’Économie et de l’Innovation du Québec, and U.S. Department of Energy, Office of Science, National Quantum Information Science Research Centers, Quantum Systems Accelerator. M.M. acknowledges support by the U.S. Army Research Office grant W911NF-16-1-0070.
The views and conclusions contained herein are those of the authors and should not be interpreted as necessarily representing the official policies or endorsements, either expressed or implied, of the ODNI, IARPA, or the U.S. Government.

\section*{Author Contributions}
S.K., N.L. and A.R. planned the experiments, S.K. and N.L performed the main experiment, and S.K. and N.L. analyzed the data.
F.S., A.R. and C.K.A. designed the device, and S.K., A.R. and G.J.N. fabricated the device.
N.L., C.H. and S.L. developed the experimental software framework and A.R., C.H., N.L., S.K. and S.L. developed control and calibration software routines.
A.R., J.H., S.K. and C.H. designed and built elements of the room-temperature setup, and S.K., A.R., C.H., S.L., N.L. and F.S. maintained the experimental setup.
S.K., N.L., A.R., C.H., S.L. and C.K.A. characterized and calibrated the device and the experimental setup.
E.G., A.D.P. and C.L. performed the numerical simulations.
M.M. provided guidance on logical qubit evaluation methodology aspects.
S.K., N.L., A.R., C.H. and S.L. prepared the figures for the manuscript and S.K., N.L., A.R., C.E. and A.W. wrote the manuscript with inputs from all co-authors.
A.B., C.E. and A.W. supervised the work.

\section*{Competing interests}
The authors declare no competing interests.

\section*{Methods}
\subsection{Controlled-phase Gates} \label{methods_sec:cz_gates}
We realize the necessary two-qubit controlled-phase (CZ) gates by tuning adjacent pairs of data (D$j$) and auxiliary qubits (X$i$, Z$i$) into resonance \cite{Strauch2003, DiCarlo2010, Negirneac2021} using individual flux lines implemented with coplanar waveguides (green in  Fig.~\ref{fig:device}b) shorted near the SQUID loop of each qubit. We fabricated all qubits with asymmetric superconducting quantum interference devices (SQUIDs) \cite{Strand2013,Hutchings2017} to allow for data qubits to idle at their minimum and auxiliary qubits at their maximum frequencies, at which the qubits are to first-order insensitive to flux noise. Data qubits are designed with idle frequencies $3.7-4.1~$GHz in a low frequency band and auxiliary qubits with idle frequencies $5.9-6.3~$GHz in a high frequency band, see red and blue/green dots, respectively, in Fig.~\ref{fig:device}c and \cref{app:device_params}.

We implement \CZs{} by tuning both data and auxiliary qubits to an intermediate interaction frequency $\omega_{\rm int}$ and $\omega_{\rm int}-\alpha$, respectively, with $\omega_{\rm int}/2\pi$ ranging from $4.4$ to $5.6~$GHz (\cref{app:two_qubit_gates}). The qubit anharmonicity $\alpha\sim -0.17\,$GHz is designed to be small to minimize residual qubit/qubit interactions~\cite{Krinner2020}. We make use of net-zero flux pulses \cite{Negirneac2021}, which reduce both the detrimental effect of low-frequency flux noise on qubit coherence and the impact of non-idealities in the transfer function of the flux lines on gate fidelities.
Given the large designed detuning of $\sim 2$~GHz between the data and auxiliary qubits at their idle frequencies, we calculate residual-ZZ interaction strengths between qubits lower than $\alpha_{\textrm{zz}}/2\pi\sim 8~$kHz \cite{Krinner2020}. It is only during two-qubit gate execution that $\alpha_{\textrm{zz}}$ increases by a factor of approximately $2$ to $25$, depending on the interaction frequency $\omega_{\rm int}$, which we partially mitigate using echo pulses. The coupling strength between auxiliary qubits and data qubits at the interaction point is about $J/2\pi\sim 7~$MHz.

\subsection{Qubit-readout Architecture} \label{methods_sec:readout}
Each qubit is coupled to a resonant pair of readout resonator and Purcell filter (red- and blue $\lambda/4$ coplanar waveguide resonators in Fig.~\ref{fig:device}b). Moreover, each readout resonator is coupled strongly to the qubit ($g/2\pi\sim 169\,$MHz for auxiliary qubits and $g/2\pi\sim 252\,$MHz for data qubits) and has a large effective bandwidth ($\kappa_{\rm eff}/2\pi\sim 10\,$MHz) to enable fast, high-fidelity readout \cite{Walter2017}. The individual Purcell filters both maintain high qubit coherence, despite the large coupling and bandwidth of the readout resonators, and reduce undesired readout crosstalk (\cref{app:crosstalk}) between qubits which are in close proximity or have similar frequencies \cite{Heinsoo2018}. This is particularly important for the simultaneous frequency-multiplexed readout of groups of four or five qubits using joint feed lines (purple coplanar waveguides in Fig.~\ref{fig:device}b). The readout resonator frequencies are separated by about 200~MHz within each feed line and occupy a frequency band extending from 6.8 to 7.6~GHz (purple points in Fig.~\ref{fig:device}c).

We read out the states of all qubits dispersively by applying frequency-multiplexed Gaussian filtered microwave pulses of duration $200-300~$ns to all four feed lines. We integrate the transmitted signals in a heterodyne detection scheme for a duration of $400$~ns (\cref{app:readout_characterization}). Auxiliary qubits are read out near their idle frequencies while data qubits are read out at a flux-tuned qubit frequency of $\sim 5$~GHz reducing the data qubit-readout resonator detuning \cite{Sank2016} and thus enhancing the dispersive coupling and the readout fidelity \cite{Wallraff2005, Walter2017}.

\subsection{Leakage Detection} \label{methods_sec:leakage}
We make use of a leakage detection scheme based on three-state readout which allows us, in postprocessing, to reject those sequences in which any of the qubits were measured in a leakage state, see \cref{app:readout_characterization}.
In our CZ gate scheme we make use of the second excited state $|2\rangle$ of the auxiliary qubits rather than the one of the data qubits to mediate the interaction which minimizes data qubit leakage.
Performing three-state readout of the data qubits after the final error correction cycle, we reject experimental runs for which data qubit leakage was detected. The rejected fraction per qubit and per cycle amounts to $0.0017(2)$. In addition there is a cycle-independent rejection probability of about $0.01$ per qubit due to false positives caused by readout-error.
In addition, we detect, if any of the eight auxiliary qubits has leaked to the $|2\rangle$ state in any of the $n$ cycles, using the same three-state readout, and find an average rejection probability of $0.0094(4)$ per qubit per cycle.
In total, this leads to a rejected data fraction per cycle of 8\,\%.
\section*{Supplementary Information}
\begin{appendix}

\section{Device fabrication} \label{app:device_fabrication}
To fabricate the 17-qubit surface code device, we pattern transmon qubit islands, couplers, resonators, and control lines into a 150\,nm-thin niobium film sputtered onto a high-resistivity silicon substrate using photolithography and reactive ion etching.
To establish a well-connected ground plane at microwave frequencies and realize cross-overs for coplanar waveguides, we fabricate aluminum-titanium-aluminum trilayer airbridges onto the device using a two-layer resist photolithography process with reflow.
We fabricate aluminium-based Josephson junctions of the transmon qubits using electron-beam lithography (EBL) and shadow evaporation, and establish electrical contact between the junction metal and the niobium base-layer film using aluminum bandages~\cite{Dunsworth2017} fabricated in a second EBL and evaporation run.

\section{Device parameters and performance} \label{app:device_params}
We characterize the performance and properties of each qubit on the device using spectroscopy and standard time-domain methods, see \cref{tab:qubit_measured_parameters}.
Note that some parameters (marked with the superscript 'a' in \cref{tab:qubit_measured_parameters}) were extracted from measurements in a separate cooldown of the same device.

We realize single-qubit gates with 40-ns-duration microwave DRAG pulses~\cite{Motzoi2009} with a Gaussian envelope which has a standard deviation of $\sigma=\ns{10}$ and is truncated at $\pm 2\sigma$.

We realize two-qubit gates with net-zero flux pulses~\cite{Rol2019, Negirneac2021} with an average duration of \ns{68}, see~\cref{app:two_qubit_gates} for a detailed description of the two-qubit gates implementation. We add \ns{15} buffers before and after each flux-pulse resulting in a total average two-qubit gate duration of \ns{98}.

\begin{figure}[t] 
\centering
\includegraphics{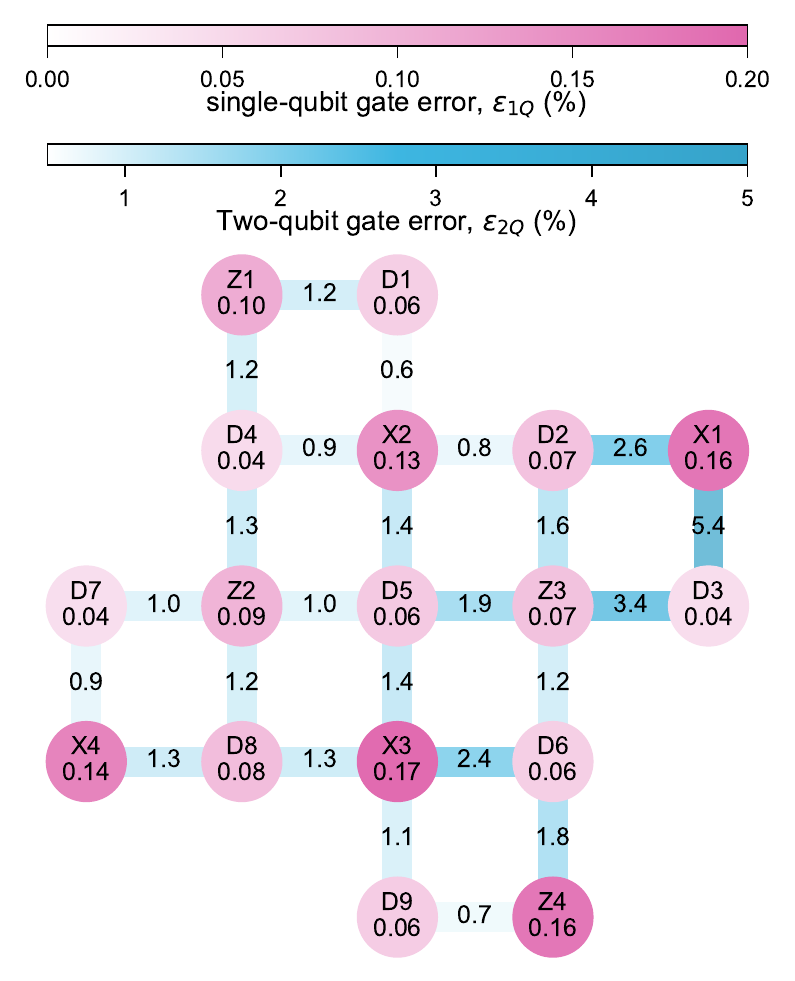}
\caption{Single- and two-qubit gate errors as characterized by randomized benchmarking and interleaved randomized benchmarking, respectively. The average measured single-qubit (two-qubit) gate error across the device is 0.09(4)\,\% [1.5(1.0)\,\%].}
\label{fig:gate_performance}
\end{figure}

We benchmark the performance of single- and two-qubit gate operations across the entire device using randomized benchmarking~\cite{Magesan2011, Epstein2014} (single-qubit gates) and interleaved randomized benchmarking~\cite{Magesan2012, Corcoles2012, Barends2014} (two-qubit gates), and display the resulting errors in \cref{fig:gate_performance}.
\CZs{} that are executed simultaneously in the quantum error correction cycle (see \cref{fig:stabilizers_state_init}a and \cref{fig:pulse_sequence}) are also calibrated and benchmarked simultaneously.

Single-qubit gates have a mean error of 0.09(4)~\%, which is in good agreement with the mean error of 0.08~\% obtained from master-equation simulations taking only decoherence into account, see \cref{app:numerical_simulations}. Two-qubit gates have an average gate error of 1.5(1.0)~\%, and a best (worst) gate error of 0.6~\% (5.4~\%).
Two-qubit gates used to realize the stabilizer \sx{1} display larger errors, which we attribute to the low coherence times of the auxiliary qubit X1, and to the interaction with a microscopic defect (\cref{app:two_qubit_gates}) with a frequency close to the idle-frequency of qubit X1.

\begin{table*}[]
    \centering
    \caption{Qubit parameters, coherence properties and single-qubit performance for the nine data qubits (top) and the eight auxiliary qubits (bottom). We also provide, for relevant quantities, the averaged value across the device in column $\overline{\mathrm{Q}}$.}
\begin{tabular}{lrrrrrrrrr}
\toprule
Parameter &       D1 &       D2 &       D3 &       D4 &       D5 &       D6 &       D7 &       D8 &       D9  \\
\midrule
Qubit idle frequency, $\omega_Q/2\pi$ (GHz)					        	        & 3.885 & 3.994 & 3.952 & 3.878 & 3.895 & 3.74 & 4.056 & 3.993 & 4.143 \\
Qubit anharmonicity, $\alpha/2\pi$ (MHz)							                    & -184 & -184 & -183 & -184 & -186 & -184 & -181 & -183 & -181 \\
Lifetime, $T_1$ (\si{\micro s})							                    & 29.1 & 33.0 & 65.5 & 59.3 & 32.2 & 60.2 & 36.0 & 33.1 & 25.9 \\
Ramsey decay time, $T_2^*$  (\si{\micro s})				        	        & 35.0 & 13.5 & 33.9 & 74.3 & 46.1 & 78.1 & 74.8 & 41.4 & 31.7 \\
Echo decay time, $T_2^\mathrm{e}$   (\si{\micro s})		        		    & 46.2 & 52.2 & 49.3 & 75.5 & 56.1 & 89.3 & 72.5 & 59.1 & 36.3 \\
Single-qubit RB error\footnotemark[1], $\epsilon_{\mathrm{1Q}}$ (\%)	    & 0.06 & 0.07 & 0.04 & 0.04 & 0.06 & 0.06 & 0.04 & 0.08 & 0.06 \\
Readout frequency, $\omega_\mathrm{RO}/2\pi$ (GHz)				        	    & 6.769 & 6.979 & 6.88 & 7.12 & 7.18 & 7.032 & 6.91 & 7.075 & 6.868\\
Qb. freq. during RO, $\omega_Q'/2\pi$ (GHz)					            	    & 5.321 & 4.75 & 5.275 & 4.25 & 4.42 & 5.13 & 4.395 & 3.993 & 5.0 \\
Dispersive shift\footnotemark[1], $\chi/2\pi$ (MHz)			        	    & -2.0 & -2.2 & -1.9 & -1.6 & -1.5 & -1.6 & -2.3 & -2.0 & -2.4 \\
Disp. shift during RO, $\chi'/2\pi$ (MHz)				            	    & -7.4 & -3.9 & -6.0 & -2.0 & -2.1 & -4.7 & -3.0 & -2.0 & -4.9 \\
Readout linewidth\footnotemark[1], $\kappa_{\mathrm{eff}}/2\pi$ (MHz)	    & 6.3 & 7.2 & 8.3 & 8.7 & 3.5 & 9.0 & 8.6 & 8.2 & 8.9 \\
Qubit-RO res. coupling\footnotemark[1], $g_{\mathrm{Q,RR}}/2\pi$ (MHz)	& 244 & 269 & 241 & 241 & 238 & 244 & 267 & 265 & 260 \\
Two-state readout error, $\epsilon_{\mathrm{RO}}^{(2)}$ (\%)			        & 0.7 & 0.5 & 0.5 & 0.6 & 0.8 & 0.5 & 2.3 & 1.3 & 0.5 \\
Three-state readout error, $\epsilon_{\mathrm{RO}}^{(3)}$ (\%)			        & 4.4 & 1.0 & 5.9 & 2.1 & 1.6 & 1.1 & 3.3 & 2.1 & 1.0 \\
Thermal population, $P_\mathrm{th}$ (\%)					                & 2.1 & 0.7 & 1.5 & 3.8 & 2.7 & 2.2 & 0.0 & 0.8 & 1.6 \\
\midrule
Parameter &       X1 &       X2 &       X3 &       X4 &       Z1 &       Z2 &       Z3 &       Z4 &  $\mathbf{\overline{Q}}$ \\
\midrule
Qubit idle frequency, $\omega_Q/2\pi$ (GHz)					        	        & 6.097 & 5.885 & 6.022 & 6.049 & 6.328 & 6.192 & 5.956 & 6.037 &\textbf{ -}\\
Qubit anharmonicity, $\alpha/2\pi$ (MHz)							                    & -170 & -174 & -170 & -170 & -163 & -168 & -171 & -170 & \textbf{-177}\\
Lifetime, $T_1$ (\si{\micro s})							                    & 12.4 & 17.4 & 18.5 & 12.6 & 17.0 & 42.7 & 29.7 & 27.5 & \textbf{32.5}\\
Ramsey decay time, $T_2^*$  (\si{\micro s})				        	        & 5.6 & 14.3 & 20.7 & 28.9 & 29.3 & 33.1 & 48.0 & 28.9 & \textbf{37.5}\\
Echo decay time, $T_2^\mathrm{e}$   (\si{\micro s})		        		    & 15.8 & 33.0 & 16.1 & 33.1 & 31.5 & 26.0 & 53.6 & 53.6 & \textbf{47.0}\\
Single-qubit RB error\footnotemark[1], $\epsilon_{\mathrm{1Q}}$ (\%)	    & 0.16 & 0.13 & 0.17 & 0.14 & 0.1 & 0.09 & 0.07 & 0.16 & \textbf{0.09}\\
Readout frequency, $\omega_\mathrm{RO/2\pi}$ (GHz)				        	    & 7.372 & 7.554 & 7.258 & 7.461 & 7.316 & 7.502 & 7.2 & 7.412 & \textbf{-}\\
Qb. freq. during RO, $\omega_Q'/2\pi$ (GHz)					            	    & 5.9 & 5.885 & 6.022 & 6.049 & 6.328 & 6.191 & 5.956 & 5.687 & \textbf{-}\\
Dispersive shift\footnotemark[1], $\chi/2\pi$ (MHz)			        	    & -2.8 & -1.9 & -3.2 & -2.6 & -4.7 & -2.9 & -3.2 & -2.8 &\textbf{ -2.4}\\
Disp. shift during RO, $\chi'/2\pi$ (MHz)				            	    & -2.1 & -1.9 & -3.2 & -2.6 & -4.7 & -2.9 & -3.2 & -2.8 & \textbf{-3.5}\\
Readout linewidth\footnotemark[1], $\kappa_{\mathrm{eff}}/2\pi$ (MHz)	    & 15.1 & 20.1 & 13.0 & 11.0 & 11.2 & 12.2 & 14.3 & 10.0 & \textbf{10.3}\\
Qubit-RO res. coupling\footnotemark[1], $g_{\mathrm{Q-RR}}/2\pi$ (MHz)	& 167 & 168 & 167 & 168 & 171 & 170 & 167 & 171 & \textbf{213}\\
Two-state readout error, $\epsilon_{\mathrm{RO}}^{(2)}$ (\%)			        & 2.7 & 0.7 & 0.8 & 0.8 & 1.3 & 0.4 & 0.4 & 0.5 & \textbf{0.9}\\
Three-state readout error, $\epsilon_{\mathrm{RO}}^{(3)}$ (\%)			        & 3.9 & 2.0 & 1.9 & 1.6 & 2.2 & 1.2 & 0.9 & 1.1 & \textbf{2.2}\\
Thermal population, $P_\mathrm{th}$ (\%)					                & 0.2 & 0.2 & 0.5 & 0.3 & 0.6 & 0.5 & 0.2 & 0.9 & \textbf{1.1} \\
\bottomrule

    \end{tabular}
    \label{tab:qubit_measured_parameters}
	\footnotetext{\label{footnote:diff_cooldown}Measured in a different cooldown.}
\end{table*}

\section{Two-qubit gates} \label{app:two_qubit_gates}
Two-qubit gates are essential building blocks of the stabilizer circuits of the surface code. Here, we describe our implementation of \CZs, discuss the constraints and considerations for gate parallelization and gate ordering during the error correction cycle, and report on coherence properties during the gate execution.

\subsection{\CZ{} implementation} \label{app:two_qubit_gates:implementation}
We realize two-qubit controlled-phase (C$Z$) gates \cite{Strauch2003, DiCarlo2010} by flux-tuning two neighboring qubits to an interaction frequency between the idle frequencies of the two qubits. We use a net-zero pulse shape to protect the qubits from low-frequency flux noise and to avoid the buildup of long-timescale distortions in the flux-control lines \cite{Rol2019, Negirneac2021}.

\begin{figure}[t] 
\centering
\includegraphics{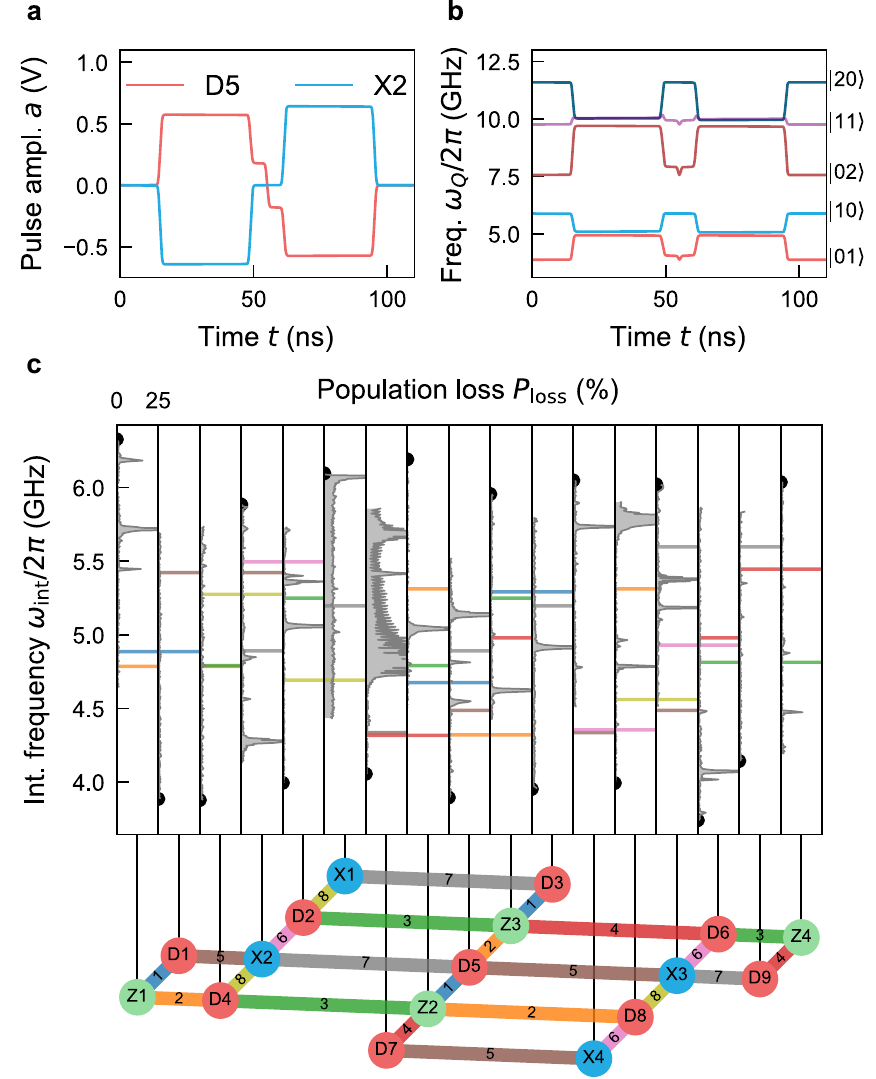}
\caption{ Realization of two-qubit gates.
	\textbf{a}. Example of a net-zero pulse shapes before compensating for flux line distortions that are applied simultaneously on a data qubit (D5) and an auxiliary qubit (X2) to implement a \CZ{} between the two. Note that the duration of the transition section has been increased from \ns{2.5} to \ns{12.5} for better visibility in this figure.
	\textbf{b}. Calculated evolution of the eigenenergies for the pulse shapes shown in panel \textbf{a}.
	\textbf{c}. Frequency arrangement chosen for the operation of the device. Measured population loss (gray areas) when flux-tuning qubits from their idle frequencies (black dots) to $\omega_{\mathrm{int}}$ for a duration of \ns{100}. The chosen two-qubit gate interaction frequencies are displayed as colored lines corresponding to the color code used in the conceptual device representation at the bottom of the panel, which indicates two-qubit pairs between which gates are executed. Gates sharing the same color are executed simultaneously, in the time step indicated by the label extending between pairs of qubits
}
\label{fig:interaction_frequencies}
\end{figure}

The flux pulse applied to each qubit consists of two Gaussian-filtered, approximately $30$-ns-long sections, which are interleaved with a 2.5-ns-long net-zero transition section, see \cref{fig:interaction_frequencies}a for an example waveform.
The amplitudes of the longer sections bring the $\ket{20}$-transition frequency into resonance with the the $\ket{11}$-transition frequency, see \cref{fig:interaction_frequencies}b.
We calibrate the duration of these sections to achieve full population recovery into the $\ket{11}$ state.

To achieve a $\pi$-controlled-phase rotation, we calibrate the pulse amplitude of the transition section for the data qubit, which effectively controls the phase acquired by the $\ket{20}$ state relative to the $\ket{11}$ state.

We select interaction frequencies which allow for the parallel execution of three \CZs{} in each of the eight time steps of the quantum error correction cycle, as indicated by the color and label of the qubit-qubit couplers in \cref{fig:interaction_frequencies}c.
The interaction frequencies for a single time step (colored horizontal lines of same color) are chosen to avoid crossing neighboring qubits on which gates are executed in the same time step.
In addition, the interaction frequencies in a given CZ gate time step are distributed over the available $\sim 2\,$GHz frequency range to minimize residual-ZZ couplings between neighboring qubits performing CZ gates in parallel. The overall, calculated residual-ZZ coupling during the eight CZ gate time steps, averaged over all 24 qubit pairs and over the eight time steps, amounts to $\alpha_{\rm ZZ}/2\pi = 27(37)\,$kHz, which is about a factor three larger than with all qubits biased at their idle frequency.

Finally, the interaction frequencies are chosen to minimize population loss due to interaction with the strongly coupled microscopic defects of our device. To characterize the defect-mode distribution, we determine the frequency-dependent population loss for each qubit by measuring the remaining excited-state population after applying Gaussian-filtered square flux pulse that tunes the qubit frequency from its idle frequency (black semi-circle) to $\omega_{\mathrm{int}}$ for a duration of $t_\mathrm{int} = \ns{100}$, see the gray filled areas in \cref{fig:interaction_frequencies}c.
For most qubits, we observe a constant background population loss of $\leq2\,\%$ over the entire frequency range,
with $0-3$ narrow frequency bands ($\leq 50\,$MHz) exhibiting peak population loss $\geq25\%$. We attribute these population-loss peaks to coherent interactions with defects coupled to the qubits with a strength of $g/2\pi \geq 0.8\,$MHz.
Qubits D7, D8 and X1 display broader and higher population-loss peaks, likely due to the interaction with defects coupled to the qubits with $g/2\pi$ on the order of 1-20 MHz.
For these qubits, the finite interaction during the rising and falling edge of the flux pulse leads to a population loss tail when crossing the defect.
We choose interaction frequencies for the two-qubit gates that are detuned from all defects and that avoid crossing strongly coupled defects.

In addition to a suitable selection of interaction frequencies, accurate control of the qubit frequency is essential for the realization of high-fidelity and low-leakage two-qubit gates.
However, the flux pulses reaching the sample and controlling the qubit frequency, are transformed from the programmed waveforms on the AWGs due to the transfer functions of, for instance, the high-pass filtering in the bias-tee, the frequency-dependent attenuation in the cables, and imperfections in the impedance matching of the flux line.
To characterize the flux pulse distortions at timescales ranging from $\sim\ns{50}$ to \SI{50}{\micro s},
we apply a step-like Gaussian-filtered flux pulse and resolve the resulting time dependence of the qubit frequency by identifying which drive frequency induces a transition from $\ket{0}$ to $\ket{1}$, for a varying delay $\Delta t$ after applying the step-like flux pulse.
The time-dependent qubit-frequency response is then used to calibrate infinite impulse response (IIR) filters
which invert the distortions of signals propagating along the flux line in digital preprocessing of the programmed waveforms, see above.
Moreover, to compensate for distortions on nanosecond timescales, we calibrate finite impulse response (FIR) filters using methods described in Ref.~\onlinecite{Rol2020}.

\subsection{Two-qubit gate order}\label{app:gate_sequence}
We choose the order of the two-qubit gates, executed in the \CZ{} time steps 1, 2, 3, 4 for $\sz{i}$ and in the time steps 5, 6, 7, 8 for $\sx{i}$ (Fig.~\ref{fig:interaction_frequencies}c), to provide resilience against single auxiliary-qubit errors propagating to data qubits~\cite{Horsman2012, Tomita2014a}. Simultaneously, we satisfy constraints imposed by the presence of microscopic defects near the two-qubit gate interaction frequencies (\cref{app:two_qubit_gates:implementation}).

\begin{figure}[t]
\centering
\includegraphics{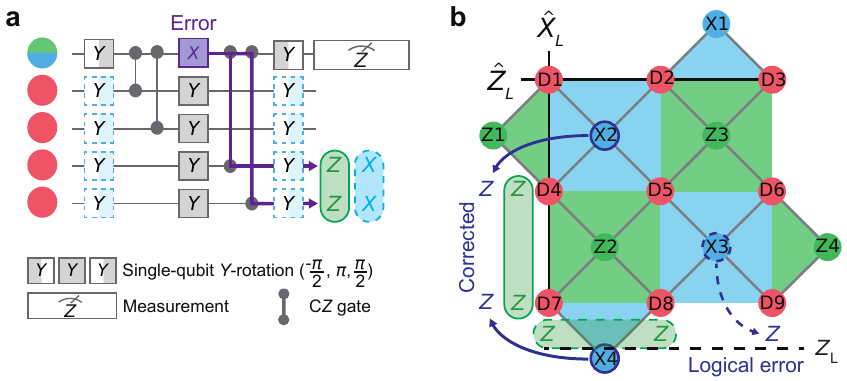}
\caption{Limiting the propagation of auxiliary-qubit errors. \textbf{a} Weight-four stabilizer circuit as discussed in the main text. A single $\hat{X}$ error on the auxiliary qubit in the middle of the circuit propagates to two data qubits (violet lines), resulting in a correlated $\hat{Z}\hat{Z}$ error (green box) for Z-type stabilizer circuits and in a correlated $\hat{X}\hat{X}$ error (blue rounded-corner box with dashed outline) for X-type stabilizer circuits. \textbf{b} 17-qubit surface code schematic with two potential correlated $\hat{Z}\hat{Z}$ errors (solid and dashed green boxes) indicated. Dark blue, solid and dashed circles around auxiliary qubits indicate the corresponding syndrome elements with value 1. Dark blue, solid and dashed arrows indicate the errors identified by the decoder, see text for details.}
\label{fig:gate_order}
\end{figure}

A single $\hat{X}$ error on an auxiliary qubit $\mathrm{Z}i$ ($\mathrm{X}i$) during a stabilizer measurement $\sz{i}$ ($\sx{i}$) results in $\hat{Z}$ ($\hat{X}$) errors on all data qubits which subsequently perform \CZs{} with that auxiliary qubit, see Fig.~\ref{fig:gate_order}a for an example.
Auxiliary qubit $\hat{Z}$ errors commute with the CZ gates and therefore do not propagate to data qubits. We only consider the middle of weight-four stabilizer gate sequences as potential times when $\hat{X}$ errors occur on the auxiliary qubits since the error propagates to two data qubits only in this case.
For all other auxiliary qubit error locations, the error effectively propagates to at most a single data qubit because data qubit errors are only relevant up to multiplication with stabilizer operators.
For instance, although an error between the first and second CZ gate physically propagates to three data qubits of the corresponding stabilizer plaquette, this three-qubit error is equivalent to a single-qubit error, of the same type, on the originally unaffected data qubit of the plaquette.

To ensure that such correlated two-qubit $\hat{Z}$ or $\hat{X}$ errors do not result in logical errors in the decoding process, we choose the last two \CZs{} of each of the $\sz{i}$ ($\sx{i}$) measurements to involve data qubits which are not aligned parallel to the data-qubit strings forming the logical operators \lopz{} (\lopx{})~\cite{Tomita2014a}, i.e. not horizontal (vertical) in \cref{fig:gate_order}b.
With this gate order, correlated two-qubit $\hat{Z}$ or $\hat{X}$ errors can be correctly identified (up to multiplication with stabilizer operators). As an example, we consider the correlated error $\hat{Z}_{\rm D4}\hat{Z}_{\rm D7}$ on D4 and D7 resulting from a single error $\hat{Z}_{\rm Z2}$ on Z2, see Fig.~\ref{fig:gate_order}b.
As a consequence, X2 and X4 show syndrome elements of 1 (dark blue, solid circles in Fig.~\ref{fig:gate_order}b). Because the stabilizers $\sx{2}$ and $\sx{4}$ do not share a data qubit, this syndrome cannot be caused by a single data-qubit error and the minimum-weight-perfect-matching decoder (\cref{app:decoding}) correctly identifies both errors, the error $\hat{Z}_{\rm D7}$ and one of the equivalent errors $\hat{Z}_{\rm D4}$ (indicated in Fig.~\ref{fig:gate_order}b with solid, dark blue arrows) or $\hat{Z}_{\rm D1}$.
On the other hand, if the gates Z2-D7 and Z2-D8 were successively performed in either the CZ gate time steps 1 and 2, or 3 and 4, the corresponding correlated error $\hat{Z}_{\rm D7}\hat{Z}_{\rm D8}$ (green rectangle with dashed outline in \cref{fig:gate_order}b) would not be correctly identified by the decoder.
In that case, only the syndrome element of X3 is 1 and as a result the decoder identifies one of the equivalent single-qubit errors $\hat{Z}_{\rm D9}$ (indicated in \cref{fig:gate_order}b with a dashed, dark blue arrow) or $\hat{Z}_{\rm D6}$, and applies the corresponding correction $\hat{Z}_{\rm D9}$ or $\hat{Z}_{\rm D6}$. However, one such false correction together with the original error $\hat{Z}_{\rm D7}\hat{Z}_{\rm D8}$ is equivalent to the application of \lopz{} and thus leads to a logical error, see black dashed line in \cref{fig:gate_order}b.

We note that data qubits participating in the last two CZ gates of a stabilizer measurement may also be aligned diagonally with respect to the data qubit string of the corresponding logical operator, and that the resulting weight-two error on the data qubits remains correctable. In particular, this is the case for $\sz{3}$ and allows us to execute the gate Z3-D5 in a different time step than the gate Z2-D7, which would otherwise result in a frequency crossing of Z2 and D5 because the gate Z2-D7 is constrained to an interaction frequency $\omega_{\rm int}/2\pi \lesssim 4.5~$GHz due to a microscopic defect on D7 (\cref{fig:interaction_frequencies}c). For completeness, we note that there is no constraint imposed on the two-qubit gate order from maintaining the commutation of neighboring stabilizers $\sz{i}$ and $\sx{i}$ \cite{Fowler2012, Tomita2014a} because we measure $\sz{i}$ and $\sx{i}$ sequentially.

\subsection{Reduced coherence at interaction frequencies}\label{app:T2_intfreq}
At the two-qubit gate interaction frequency, the sensitivity of the qubit frequency to fluctuations in flux is higher than at the idle frequency, which is first-order insensitive to fluctuations in flux. Hence, flux-noise-induced dephasing is increased during two-qubit gates.
We characterize the effective Ramsey decay time $T_2^{\ast,\mathrm{int}}$, for a given qubit $Qi$ ($Q \in \{\mathrm{D}, \mathrm{X}, \mathrm{Z}\}$ and $i=1, \dots, 9$ if $Q = \mathrm{D}$ and $i=1, \dots, 4$ otherwise), fluxed-tuned to the interaction frequency $\omega_\mathrm{int}$ with a sequence of Ramsey measurements.
Specifically, we measure the phase of a given qubit with a Ramsey experiment in which we insert a train of $N$ \ns{60}-duration net-zero flux pulses (with 15-ns-long buffers on each side) between two Ramsey $\pi/2$-pulses.
We repeat the measurement for $N=1,2,4,8,\dots,256$ pulses, extract the phase contrast for each measurement from a cosine fit, and fit the phase contrasts to a decaying exponential to extract $T_2^{\ast,\mathrm{int}}$.

\begin{figure}[t] 
\centering
\includegraphics{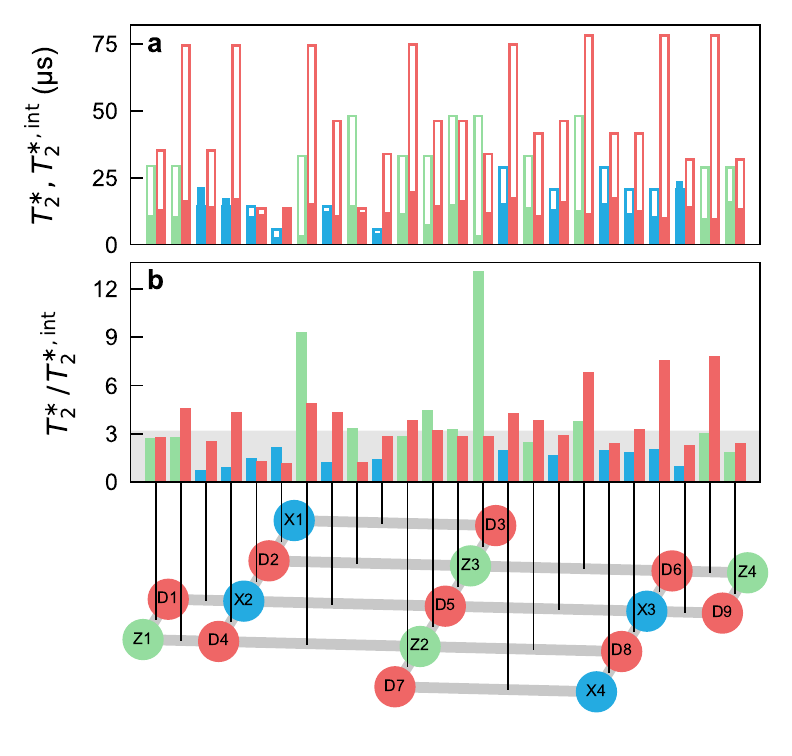}
\caption{Reduced coherence during two-qubit gate execution.~\textbf{a} Comparison between the Ramsey decay times measured at the qubit idle frequencies (wire-frames), $T_2^{\ast}$,  and at the two-qubit gate interaction frequencies (solid bars), $T_2^{\ast,\mathrm{int}}$,  for the 24 two-qubit gates of the device, with colors matching the respective qubit color in the conceptual device representation at the bottom of the figure. Each qubit appears multiple times as it is tuned to different interaction frequencies depending on which neighboring qubit it performs a gate with.~\textbf{b} Ratio of the Ramsey decay time at the qubit idle frequency, $T_2^{\ast}$, and the Ramsey decay time at the two-qubit gate interaction frequency, $T_2^{\ast,\mathrm{int}}$. The median ratio is indicated by the gray area.
}
\label{fig:t2_interaction_frequencies}
\end{figure}

We compare $T_2^{\ast,\mathrm{int}}$ to the Ramsey decay time measured at the qubit idle frequency, $T_2^{\ast}$, for each of the 24 two-qubit gates, see the wire-frame and filled bars in \cref{fig:t2_interaction_frequencies}a, respectively.
Averaged over the 24 qubit-pairs, we observe a mean $T_2^{\ast\mathrm{int}}$ of 13.1(4.2)$\,$µs at the interaction frequency, compared to a mean $T_2^{\ast}$ of 46.8(20.2)$\,$µs at the idle frequency.
As the built-in echo effect of the net-zero flux-pulse provides protection against low-frequency flux-noise, we hypothesize that the reduction in decay-time is dominated by high-frequency flux noise.
We compute the decay-time ratio $T_2^{\ast}/T_2^{\ast,\mathrm{int}}$ for each qubit-pair, see~\cref{fig:t2_interaction_frequencies}b, and extract a mean ratio of $\sim3.2$ (gray area).
This reduction in Ramsey decay time significantly affects the coherence limit of the two-qubit gates. To reproduce the characteristics of the experiment in numerical simulations, we account for this reduction by adjusting the dephasing rates for the duration of the two-qubit gates, see \cref{app:numerical_simulations} for details.

\section{Experimental setup} \label{app:experimental_setup}
\begin{figure*}[t] 
\centering
\includegraphics[width=0.98\textwidth]{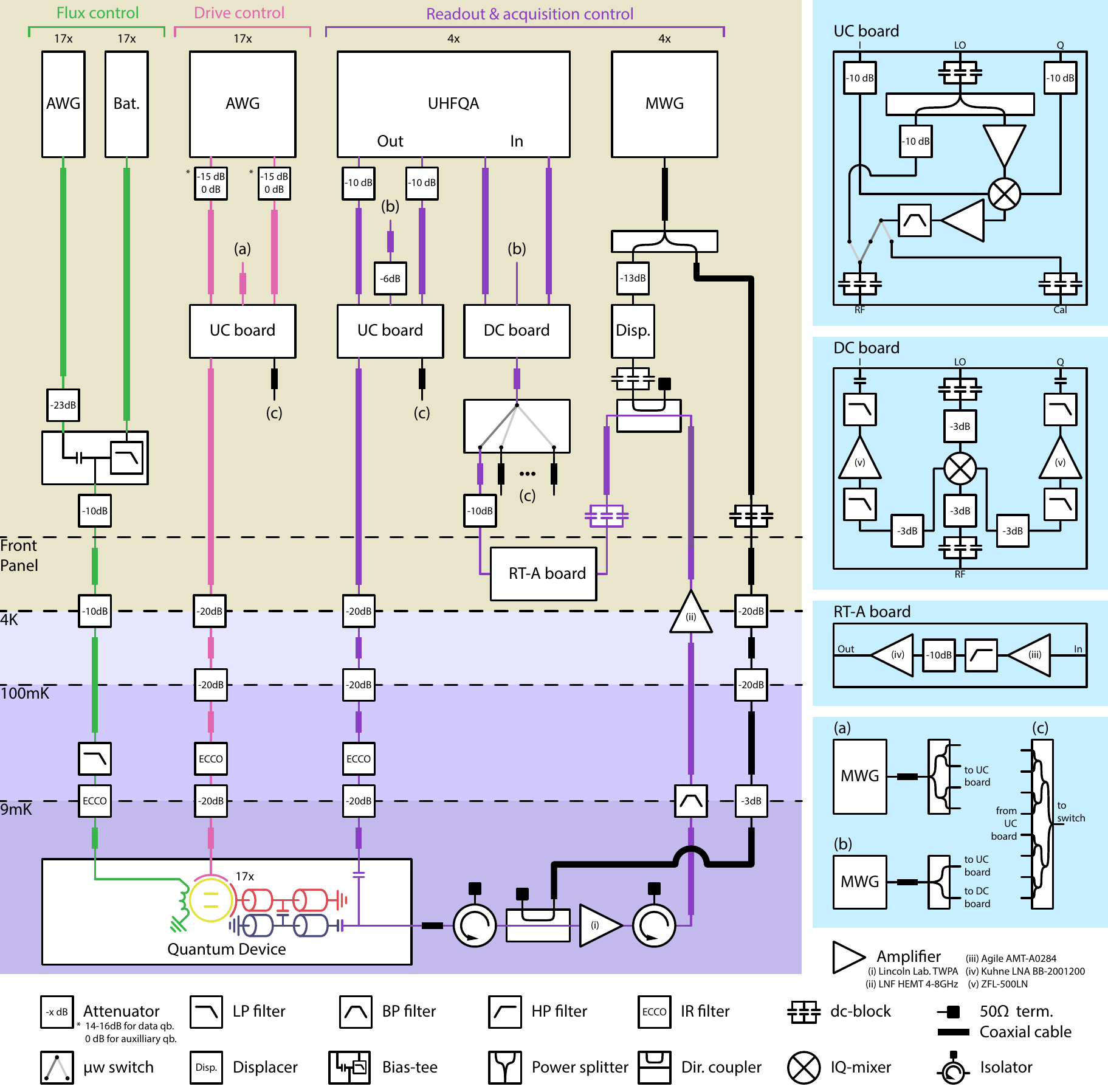}
\caption{Schematic illustration of the experimental setup. Flux lines (green), drive lines (pink), and readout lines (purple) connect room-temperature electronics to the device, schematically represented by qubits (yellow), readout resonators (red) and Purcell filters (dark blue). The background colors indicate the temperature stages of the experimental setup. We provide additional details about the components in the legend (white background) and in the text.}
\label{fig:experimental_setup}
\end{figure*}

We install the 17-qubit quantum device in a magnetically-shielded sample holder mounted at the base plate of a dilution refrigerator~\cite{Krinner2019} and connect it to the control-electronics setup located at room temperature as indicated in \cref{fig:experimental_setup}. Input and output lines for charge control (pink), flux control (green), and readout (purple), are configured with the indicated microwave components for signal conditioning.

DC voltage sources (Bat.) generate a current which passes through a series of attenuators and filters to induce a magnetic flux in the SQUID-loop of the transmon qubit and control its idle frequency.
Arbitrary waveform generators (AWG) generate voltage pulses at a sampling rate of 2.4 GSa/s to control the qubit frequencies on the nano-second timescale to implement two-qubit gates, see \cref{app:two_qubit_gates}.  The AWG signal is combined with the DC bias current using a bias-tee.

Single-qubit drive pulses are generated at an intermediate frequency in the range of 0-\SI{500}{\mega\hertz} by an AWG, and then up-converted (UC) to microwave frequencies in an analog IQ-mixer using the continuous-wave signal of a microwave generator (MWG) as a carrier.

An ultra-high frequency quantum analyzer (UHFQA) generates multiplexed-readout pulses at a sampling rate of 1.8 GSa/s. The amplification chain at each readout port of the sample consists of a wide-bandwidth near-quantum limited traveling-wave parametric amplifier (TWPA) \cite{Macklin2015}, a high-electron-mobility transistor (HEMT) amplifier and low-noise, and room-temperature amplifiers (RT-A board).
We add the TWPA pump tone to the input of the amplifier using a 20 dB directional coupler (Dir. coupler), and cancel it interferometrically at the input of the room-temperature amplifiers by combining the cryostat output signal with a phase- and amplitude-displaced pump tone that bypasses the cryostat.
After amplification, the signal is down-converted with an IQ-mixer in a down-conversion (DC) board and then both digitally demodulated and integrated using the field-programmable gate-array of the UHFQA.

\section{Crosstalk characterization and compensation} \label{app:crosstalk}

An important requirement for scaling up quantum processors is the individual and independent control of its constituents.
Here, we characterize three types of crosstalk relevant for the execution of the surface code on our 17-qubit device: microwave drive crosstalk, flux crosstalk, and measurement-induced dephasing.

\subsection{Drive crosstalk}

\begin{figure}[t] 
\centering
\includegraphics[width=\columnwidth]{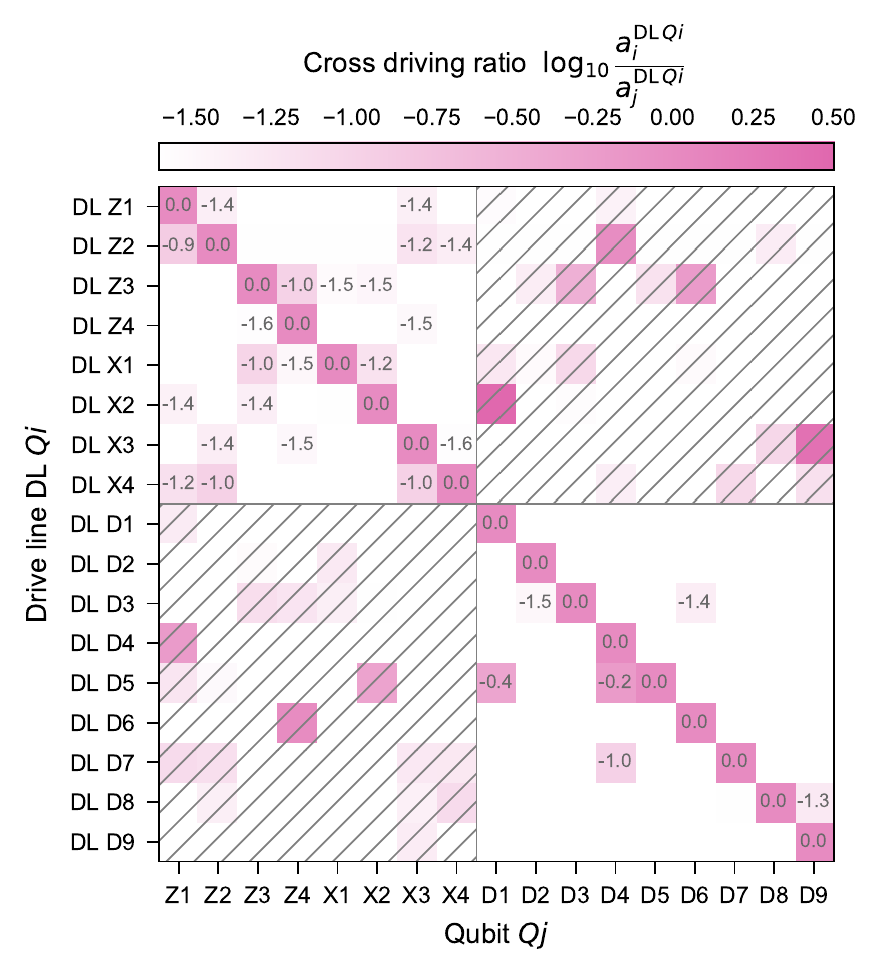}
\caption{Drive crosstalk characterization. For each drive line (DL) targeting qubit $Qi$, crosstalk is expressed as the logarithm of the ratio between the amplitude of a $\pi$-pulse on the target qubit, $a_i^{\mathrm{DL} Qi}$, and the amplitude of a $\pi$-pulse on any other qubit $j$, $a_j^{\mathrm{DL} Qi}$. The hatched gray area corresponds to cross-driving of far detuned qubits ($\sim$2 GHz), see text for details. White matrix elements without annotation correspond to DL-qubit pairs with cross-driving ratios $<-1.6$.}
\label{fig:drive_crosstalk}
\end{figure}
During the execution of multi-qubit quantum circuits, individual qubits are susceptible to off-resonant driving mediated by microwave pulses applied to drive lines designed to address other qubits.
Ensuring low cross-driving is therefore essential for the successful execution of the surface code.

We characterize the coupling rates of each drive line (DL $Qi$) to all other qubits on the device, by measuring the excited state population of qubit $Qj$ after applying a 100-ns-long Gaussian drive pulse to drive line DL $Qi$ at the frequency of qubit $Qj$. We sweep the pulse amplitude and fit a cosine function to the measured data to obtain the amplitude $a_j^{\mathrm{DL}\, Qi}$ corresponding to a $\pi$-rotation induced on qubit $Qj$. As a measure of cross-driving, we normalize each value by the $\pi$-pulse amplitude of the target qubit $a_i^{\mathrm{DL}\,Qi}$ to obtain the cross-driving ratios, of which we show the negative logarithm  $-\log_{10}\left({a_j^{\mathrm{DL} \,Qi}}/{a_i^{\mathrm{DL}\,Qi}}\right)$ in \cref{fig:drive_crosstalk}. Given the chosen pulse duration and the maximum drive rate which we can apply without compressing the up-conversion IQ-mixer, this method allows us to resolve cross-driving ratios down to about $-\log_{10}\left({a_j^{\mathrm{DL} \,Qi}}/{a_i^{\mathrm{DL}\,Qi}}\right)\approx -1.6$, or equivalently a Rabi rotation angle of 4.6 degrees. We find the off-diagonal elements of the cross-driving matrix on average 1.5 orders of magnitude smaller than the drive rate of the target qubit, which provides an upper bound for the cross-driving of $Qj$ via DL $Qi$, since pulses applied to DL $Qi$ during the execution of the surface code are detuned from the frequency of $Qj$.

Some auxiliary-data qubit pairs in physical proximity exhibit larger crosstalk, most likely mediated by the qubit-qubit coupling resonator.
However, for all these pairs, the drive line addresses a qubit in the other frequency band (see \cref{fig:device}a and \cref{fig:device}c for a visual representation of the geometry and the frequency bands) than the cross-coupled qubit (see hatched region in \cref{fig:drive_crosstalk}), which strongly suppresses the effective cross-driving during device operation. Taking both the relative detunings and the experimentally characterized cross-driving ratios into account, we estimate the gate error induced on $Qj$ resulting from applying a $\pi$-pulse on $Qi$ via DL $Qi$ and find that the gate error is smaller than 0.01\% for all auxiliary-data qubit pairs except for DL X3 - D9 and DL X2 - D1 with estimated gate errors of 0.06\% and 0.17\%, respectively.

For qubit pairs within the same frequency band, we find that the expected gate error is smaller than 0.2\% for all pairs except
for the drive line DL D5 and qubits D1 and D4, for which the expected gate errors are 19.2\% and 16.6\%, respectively. We attribute the stronger cross-coupling for these elements to the crossover between DL D5 and the qubit-qubit couplers connecting Z1-D1 and D4-X2 (see \cref{fig:device}b).
As crosstalk on that scale would significantly impact the execution of the surface code, we compensate the cross-driving of DL D5 on D1 and D4 with an interferometric cancellation drive pulse which has opposite amplitude but is otherwise identical to the pulse arriving at D1 via DL D5.

Note that this characterization measurement was performed in a separate cooldown of the same device.

\subsection{Flux crosstalk}

During the execution of a two-qubit gate, the transition-frequencies of both qubits are tuned to an intermediate interaction frequency (\cref{app:two_qubit_gates}), at which the sensitivity of the qubit-frequency to flux is increased. Consequently, the parallel execution of two-qubit gates requires a careful flux crosstalk characterization and compensation.

\begin{figure*}[t] 
\centering
\includegraphics{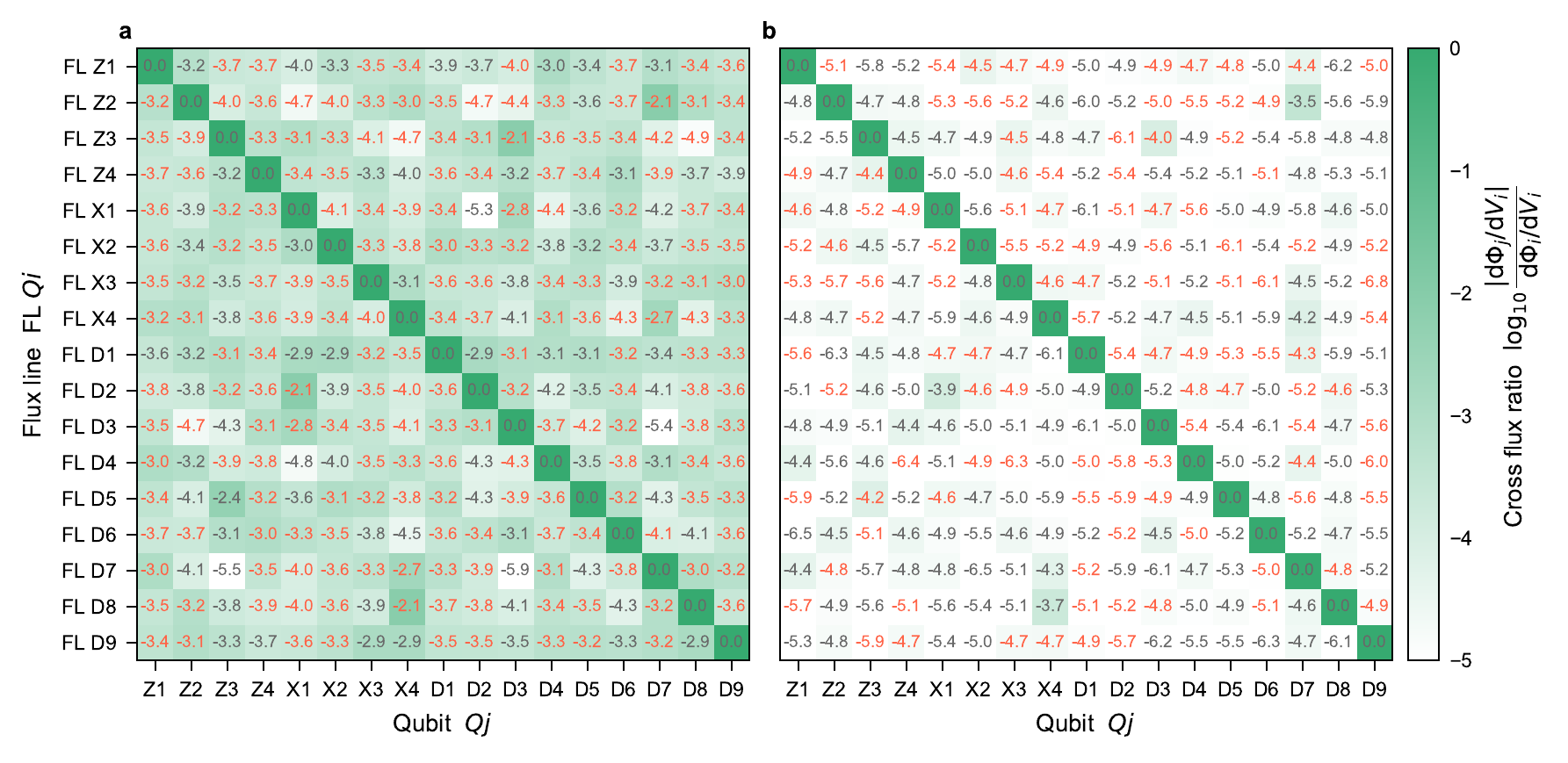}
\caption{Flux crosstalk characterization and compensation. \textbf{a}~Measured flux crosstalk matrix \fcmtx{} normalized by its diagonal, for a \ns{60} net-zero flux pulse and without crosstalk compensation (see text for details). \textbf{b}~Measured flux crosstalk matrix with a single round of flux crosstalk compensation calibrated based on the matrix shown in \textbf{a}. Red numbers indicate a negative sign of $\mathrm{d}\Phi_j/\mathrm{d}V_i$.}
\label{fig:flux_crosstalk}
\end{figure*}

We characterize the coupling of each flux line on the device to all qubits using a sequence of Ramsey experiments.
In particular, we measure the effect of the flux line with target qubit $Qi$ (FL $Qi$) on any qubit $Qj$ by sweeping the amplitude $V_i$ of a voltage pulse on FL $Qi$ and measuring the phase \fcdeltaphase{} induced on $Qj$ in a Ramsey experiment.
During the experiment, $Qj$ is flux-tuned away from its idle frequency to increase its flux sensitivity and thereby increase the signal-to-noise ratio of the measured phase.
We convert the measured phase to frequency of $Qj$, $\fcdeltafreq=\fcdeltaphase/\tau_{\rm fp}$, with $\tau_{\rm fp}=60\,$ns the duration of the flux pulse applied on FL $Qi$, which is close to the mean two-qubit gate duration on our device.
Finally, we convert the frequency $\fcdeltafreq(V_i)$ to a flux $\fcdeltaflux(V_i)$ dependent on the applied voltage $V_i$ and fit a linear function to it, whose slope ${\rm d}\Phi_j/{\rm d}V_i$ corresponds to an element of the flux crosstalk matrix \fcmtx.
Repeating this procedure for all flux lines and all qubits, we obtain the flux crosstalk matrix.
For each flux line FL $Qi$, we normalize all crosstalk elements with respect to the targeted coupling on $Qi$, i.e. ${\rm d}\Phi_i/{\rm d}V_i$, to obtain the cross-flux ratio, which we show on a base-ten logarithmic scale in \cref{fig:flux_crosstalk}a.

We find that crosstalk (off-diagonal elements) is on average about three orders of magnitude smaller than the target couplings (diagonal elements).  The total induced flux $\vec{\Phi} = (\Phi_1, \dots, \Phi_{17})^\mathrm{T}$ in the SQUID loops when applying voltage pulses with amplitude $\vec{V}= (V_1, \dots, V_{17})^\mathrm{T} $  is given by
\begin{equation}
\vec{\Phi} = \fcmtx\,\vec{V}
\end{equation}
Hence, to induce a target flux vector $\vec{\Phi}'$, we program the amplitudes
\begin{equation}
\vec{V}'= \fcmtx^{-1}\,\vec{\Phi}'
\end{equation}
which compensates for the flux line crosstalk between all qubits on the device.

To verify the effectiveness of our compensation scheme, we repeat the crosstalk characterization measurement with activated flux crosstalk compensation, which yields $\tilde{\mathbf{C}}$, see \cref{fig:flux_crosstalk}b. The resulting off-diagonal elements of $\tilde{\mathbf{C}}$ are suppressed by two additional orders of magnitude.
To further reduce the flux crosstalk in our experiments, we apply the crosstalk compensation method recursively by using $\tilde{\mathbf{C}}^{-1}$ as a second compensation matrix.

\subsection{Measurement-induced dephasing}
A successful implementation of the surface code requires repeated readout of a subset of auxiliary qubits without disturbing the state of any other qubits. However, in the presence of finite readout crosstalk, a readout pulse on an auxiliary qubit can off-resonantly excite the readout resonator of another qubit, thereby leading to dephasing and/or coherent phase rotations on that other qubit.

To characterize this effect, we sweep the amplitude of the readout pulse of an auxiliary qubit \midqbm{} while measuring the phase of another qubit \midqbd{} in a Ramsey experiment~\cite{Heinsoo2018}.
We fit the Ramsey-fringes contrast to a Gaussian model and the phase deviation to a quadratic model, from which we extract the measurement-induced dephasing rate $\Gamma_{ij}$, and the coherent phase rotation $\Delta\phi_{ij}$ for the untargeted qubit \midqbd{} when reading out \midqbm. The additional dephasing rate can be related to a phase-flip probability $P_\phi^{ij} = [1 - \exp(-\Gamma_{ij} \tau_{\textrm{RO}})]/2$, where $\tau_{\textrm{RO}}$ is the readout pulse duration.

\begin{figure}[t] 
\centering
\includegraphics{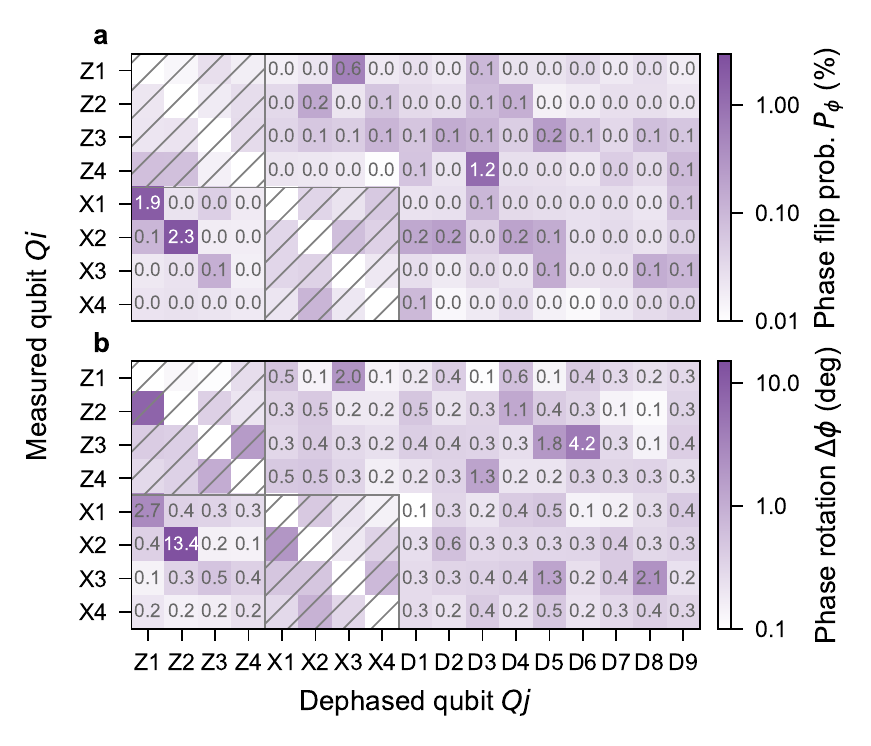}
\caption{Measurement-induced dephasing. \textbf{a}~Phase flip probability and \textbf{b} coherent phase rotation when measuring auxiliary qubits, on any other qubits. The hatched gray area corresponds to auxiliary qubits which are read out simultaneously in the surface-code experiment (see text for details).}
\label{fig:measurement_induced_dephasing}
\end{figure}

We characterize the measurement-induced dephasing of all auxiliary qubits on any other qubit, see \cref{fig:measurement_induced_dephasing}. Note that for this measurement, the qubit idle frequency of X1 was changed from 6.097$\,$GHz to 4.429$\,$GHz to avoid a microscopic two-level defect whose frequency drifted towards X1's idle frequency.

On average, we observe a phase-flip probability of $0.09\,\%$ and a coherent phase rotation of $0.6^{\circ}$. In our implementation of the surface code, all Z-type (X-type) auxiliary qubits are measured simultaneously, and therefore their mutual measurement-induced dephasing is not a concern (hatched region in \cref{fig:measurement_induced_dephasing}). We observe the largest dephasing ($P_\phi = 2.3\,\%$ and $\Delta\phi = 13.4^\circ$) on auxiliary qubit Z2, when reading out auxiliary qubit X2, which we attribute to cross-driving of the readout resonator of Z2 by the readout signal targeting X2. This dephasing could partially explain the higher mean syndrome element $\sigma_m^{\mathrm{Z}2}$, compared to the other mean syndrome elements (see \cref{fig:stabilizers_state_init}c).

In future work, we expect that we could correct for the coherent phase rotations by utilizing virtual-$Z$ rotations of equal magnitude and opposite sign.

\section{Readout characterization} \label{app:readout_characterization}

\subsection{Three-state readout}
\begin{figure}[t] 
\centering
\includegraphics{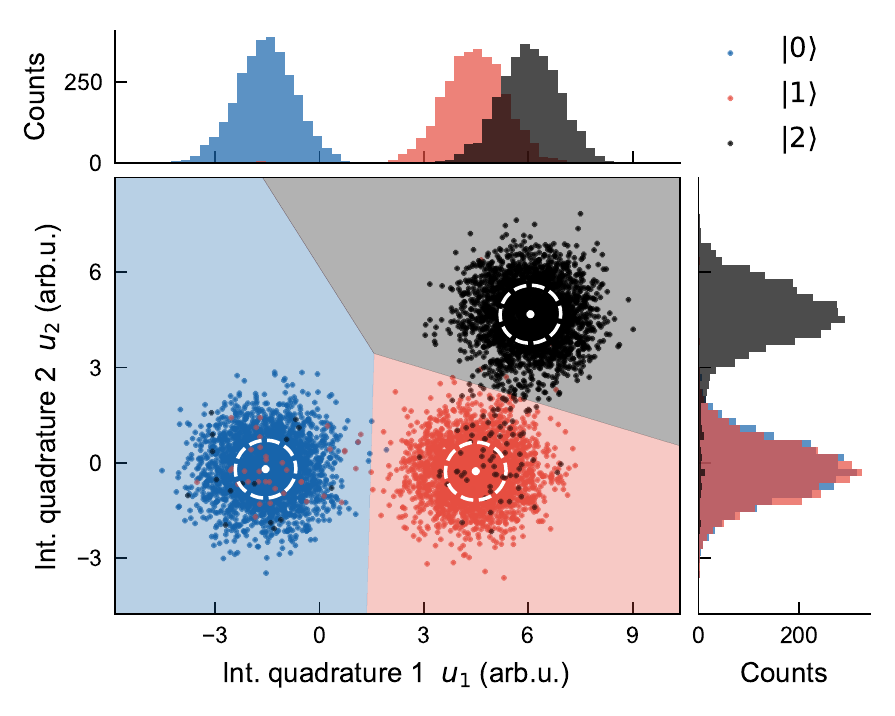}
\caption{Three-state readout characterization of auxiliary qubit X4 presented as a representative example. We show the first 3000 of the $10^5$ single-shot measurements for each of the three state preparations, after preselection. The
blue, red, and black areas delimit the regions of the
integrated quadrature plane in which a measured data point is
assigned to $\ket{0}$, $\ket{1}$ and $\ket{2}$, respectively. We indicate the mean (white dot) and the $1 \sigma$ confidence ellipse (white dashed line) of each Gaussian distribution. The marginal histogram distributions of $u_1$ and $u_2$ are shown in the top and right panel, respectively.}
\label{fig:three_level_readout}
\end{figure}

We dispersively read out the state of the transmon qubits by applying a Gaussian-filtered ($\sigma = \SI{10}{ns}$) rectangular pulse with a pulse duration of 200\,ns to auxiliary qubits and 300\,ns to data qubits, see \cref{fig:fpaRO}b for an example waveform.
To optimally distinguish between the first three states of the transmon, we integrate the complex-valued downconverted signals $s(t)$ in real-time on the UHFQA (see \cref{app:experimental_setup}) with two sets of complex-valued, 400-ns-long integration weights $w_i(t)$  \cite{Walter2017, Heinsoo2018, Kurpiers2018, Magnard2018, Lacroix2020} to obtain
\begin{equation}
u_i = \mathrm{Re}\left\{ \int_0^{t_\mathrm{int.}} s(t) w_i(t) \,\mathrm{d} t \right\}.
\end{equation}
We use the integration weights
\begin{eqnarray}\label{key}
	w_1(t) &=& s_{\ket{1}}^*(t) - s_{\ket{0}}^*(t), \\
    w_2(t) &=& s_{\ket{2}}^*(t) - s_{\ket{0}}^*(t)- \\
     && -\frac{\int w_1(t)(s_{\ket{2}}(t) - s_{\ket{0}}(t)) \,\mathrm{d}t}{\int |w_1(t)|^2 \,\mathrm{d}t} w_1(t),
\end{eqnarray}
where $s_i(t)$ is the averaged measured readout-resonator response for a qubit prepared in state $i \in \{\ket{0},\ket{1},\ket{2}\}$.

To characterize the single-shot readout, we prepare the qubit $10^5$ times in each of the three basis states ($\ket{0}$, $\ket{1}$, and $\ket{2}$), and measure the integrated resonator response yielding a pair of values $\{u_1, u_2\}$ for each experimental run. Prior to applying the state-preparation pulses we perform a pre-selection readout and reject measurements in which the qubit is not in the ground state. We estimate the thermal population for each qubit as the probability to be in $\ket{1}$ or $\ket{2}$ after this pre-selection readout, see \cref{tab:qubit_measured_parameters}.
We fit the distribution of measured $\{u_1,u_2\}$ pairs to a trimodal Gaussian mixture model, associating each qubit state to one of the Gaussian components,  see \cref{fig:three_level_readout}.
To characterize two-state readout, we use only the Gaussian components of the mixture model corresponding to the $\ket{0}$ and $\ket{1}$ state.
Based on the fitted model, we assign the individual readout outcomes to the most likely qubit state and compute the $N$-level readout error (\cref{tab:qubit_measured_parameters})
\begin{equation}
    \epsilon_{\textrm{RO}}^{(N)} = 1 - \frac{1}{N}\sum_{i = 1}^N P(i|i),
\end{equation}
where $P(i|i)$ is the probability of correctly assigning the state $i$ to a qubit prepared in the state $i$.

\subsection{Flux pulse-assisted readout}
\begin{figure}
\centering
\includegraphics{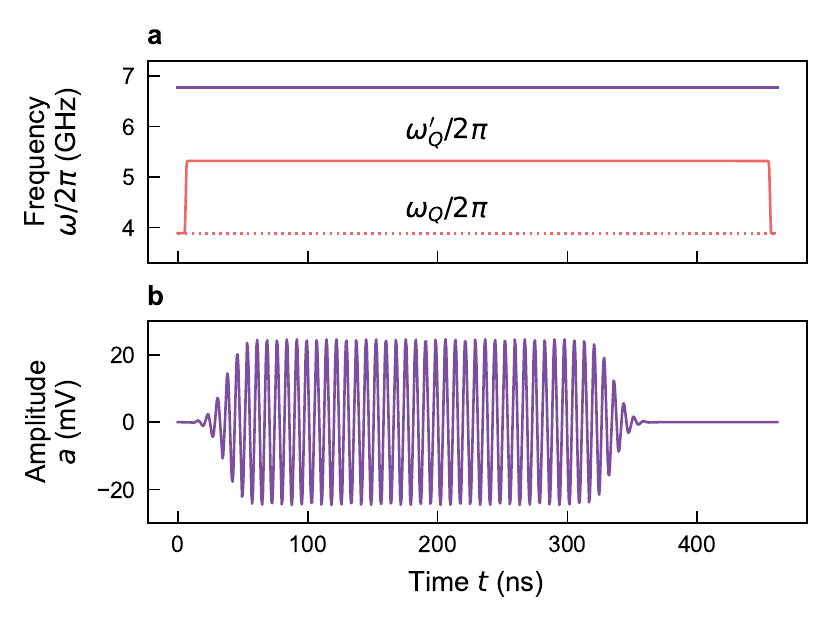}
\caption{Pulse sequence used for flux pulse-assisted readout of qubit D1. \textbf{a} The qubit transition-frequency (red solid line) is tuned from its lower sweet spot value (red dotted line), $\omega_Q \sim 4\,$GHz, to a value $\omega_Q'$ closer to the readout resonator frequency (purple line), using a fast flux pulse. \textbf{b} Intermediate frequency signal programmed on the AWG to generate the readout pulse.}
\label{fig:fpaRO}
\end{figure}

We make use of the flux-tunability of our transmon qubits to dynamically change the qubit frequency for the duration of the readout, see \cref{fig:fpaRO}a. This allows us to optimize readout parameters {\it in-situ}, such as the readout resonator-qubit detuning and the dispersive shift $\chi$, see~\cref{tab:qubit_measured_parameters} for the parameters during readout. We employ this method for all data qubits to reduce the detuning with the readout resonator during the readout, and for auxiliary qubit X1 to avoid a microscopic defect located close to its idle frequency (\cref{fig:interaction_frequencies}c).

We use Gaussian-filtered, rectangular flux pulses with short rising and falling edges ($\sigma=0.5$~ns) in order to minimize coupling to defects. The flux pulse lasts longer than the readout pulse (see also \cref{app:pulse_sequence}) because we continue integrating the readout signal while the readout resonator field is ringing down.

\section{Numerical Simulations} \label{app:numerical_simulations}
\subsection{General considerations}

In this section, we describe how the simulations of the experiment are performed. These simulations are based on the measured device characteristics including the individual qubit coherence times, readout errors (see~\cref{tab:qubit_measured_parameters}) and spurious $ZZ$ interactions.

\subsubsection{Device Hamiltonian}

The system Hamiltonian includes a flux-tunable transmon Hamiltonian per qubit mode and capacitive-coupling interactions of the form~$\hbar g_{k,k'} \hat{n}_k\hat{n}_{k'}$, where~$\hat{n}_k$ is the charge operator associated with the~$k$th qubit, and~$\hbar g_{k,k'}$ is the capacitive coupling strength between the pair of qubits~$(k,k')$. Single-qubit gates are modeled by a microwave-drive Hamiltonian of the form~$2e\hat{n}_k V_k(t)$, where~$V_k(t)$ is a time-dependent voltage. Two-qubit gates are modeled by time-dependent flux pulses~$\{\phi_k(t),\phi_{k'}(t)\}$ on both target and control qubits in the pair~$(k,k')$. Here,~$\phi_k$ is the external flux applied to the SQUID loop of qubit~$k$.

To facilitate the comparison to spectroscopy data used for gate calibration, we move to the flux-dependent basis which diagonalizes the device Hamiltonian,
resulting in an effective model of the form
\begin{align}
\begin{split}
    \hat{H}_\mathrm{eff}(t)/\hbar = & \sum_k \omega_k[\phi_k(t)] b^\dagger_k b_k + \frac{\alpha_k}{2}b^{\dagger 2}_k b^2_k  -i \Omega_k(t) (b_k - b^\dagger_k) \\
    &+ \sum_{k<k'} \xi_{k,k'}[\phi_k(t),\phi_{k'}(t)] b^\dagger_k b_k b^\dagger_{k'}b_{k'}\\
    &+ \dots
    \label{eq: effective circuit Hamiltonian}
\end{split}
\end{align}
Here,~$\omega_k[\phi_k(t)]$,~$\alpha_k$ and~$\Omega_k(t)$ correspond, respectively, to the frequency, anharmonicity and effective drive amplitude of the~$k$th qubit mode, while~$\xi_{k,k'}[\phi_k(t),\phi_{k'}(t)]$ is the strength of the cross-Kerr (or $ZZ$) interaction of the form $b^\dagger_k b_k b^\dagger_{k'}b_{k'}$ between the pair of modes~$(k,k')$. Note that the explicit dependence of the qubit frequencies and cross-Kerr interactions on the flux biases~$\phi_k$ is omitted below.
This model accounts for the change in the $ZZ$-interactions with the bare mode frequencies, and therefore with the external flux biases~$\{\phi_k(t),\phi_{k'}(t)\}$ during the two-qubit gates.
Higher-order interactions represented by the dots in~\cref{eq: effective circuit Hamiltonian} are not included in this effective model. Indeed, the impact of such nonidealities is minimized by device design and careful scheduling of the two-qubit gates, as described in \cref{app:crosstalk,app:two_qubit_gates}.

We describe relaxation and dephasing with the effective zero-temperature master equation for the system's density matrix~$\rho(t)$
\begin{align}
\begin{split}
    \dot{\rho}(t) &= -i[\hat{H}_\mathrm{eff}(t)/\hbar,\rho(t)] \\
    &+ \sum_k \gamma_{1,k} \mathcal{D}[b_k]\rho(t) \\
    & + \sum_k \gamma_{\varphi,k}[\phi_k(t)] \mathcal{D}[b^\dagger_k b_k]\rho(t),
    \label{eq: effective master equation}
\end{split}
\end{align}
where~$\gamma \mathcal{D}[\hat{o}]\bullet = \mathcal{D}[\hat{c}]\bullet = \hat{c}\bullet \hat{c}^\dagger -\{\hat{c}^\dagger\hat{c},\bullet\}/2$ for the collapse operator~$\hat{c}=\sqrt{\gamma}\hat{o}$, and ~$\gamma_{1,k}$ and~$\gamma_{\varphi,k}[\phi_k(t)]$ are, respectively, the decay and dephasing rates associated to mode~$k$. The dephasing rates of the qubits involved in two-qubit gates incorporate a flux-bias dependence~$\gamma_{\varphi,k}[\phi_k(t)]$ which is determined experimentally.

\subsubsection{Numerical solver}

To make the simulation of the 17-qubit chip numerically tractable, with each qubit modeled as a $d$-dimensional Kerr-nonlinear oscillator, we employ the method of Monte Carlo wavefunctions~\cite{Dalibard1992,Molmer1993,Dum1992} as implemented in QuTiP's \texttt{mcsolve}~\cite{Johansson2012a}. Because it evolves wavefunctions of size $d^{17}$ rather than the $d^{17}\times d^{17}$ density matrix necessary for a master-equation simulation, the memory requirements are significantly reduced in this approach with respect to the master-equation simulation.

Succinctly, the Monte Carlo method evolves a stochastic wavefunction~$|\psi(t)\rangle$ according to a non-Hermitian Hamiltonian~$H_\mathrm{nh} = H_\mathrm{eff}(t) - i\frac{\hbar}{2} \sum_l c_l^\dagger c_l$, where~$c_l$ are the collapse operators of the system's master equation. Time-evolution under $H_\mathrm{nh}$ for a time $dt$ leads to a decrease of the norm-square of the wavefunction by $dp = \sum_l dt\langle \psi(t)|c_l^\dagger c_l|\psi(t)\rangle\ll 1$. At time~$t+dt$, the wavefunction~$|\psi(t+dt)\rangle$ is renormalized with probability~$1-dp$, or subject to a single quantum jump with probability~$dp$. In the case of a jump, the wavefunction collapses to the state~$c_{l}|\psi(t)\rangle/{ \langle \psi(t)|c_{l}^\dagger c_{l}|\psi(t)\rangle}^{1/2}$ with relative probability~$\langle \psi(t)|c_{l}^\dagger c_{l}|\psi(t)\rangle/\sum_{l'} \langle \psi(t)|c_{l'}^\dagger c_{l'}|\psi(t)\rangle$.

One realization of this stochastic evolution is known as a quantum trajectory, and an advantage of this approach is that multiple such trajectories can be numerically computed in parallel \cite{Molmer1993}. With a sufficiently large number of trajectories, one can recover the solution of the master equation in~\cref{eq: effective master equation} as
\begin{equation}
    \mathbb{E}[|\psi(t\rangle\langle \psi(t)|] = \rho(t).
    \label{eq: mcsolve convergence}
\end{equation}
Expectation values of an observable~$\hat{O}$ are similarly obtained from
\begin{equation}
    \mathbb{E}[\langle\psi(t)|\hat{O}|\psi(t)\rangle] = \mathrm{tr}[\rho(t) \hat{O}].
    \label{eq: mcsolve convergence observable}
\end{equation}
We adjust the solver's error-tolerance parameters such
that~\cref{eq: mcsolve convergence} holds numerically (see~\texttt{qutip.Options}). To do so,
we determine the appropriate solver parameters by comparing the result of~\texttt{qutip.mcsolve} to that produced by the complete master-equation solver~\texttt{qutip.mesolve} for systems of up to seven qubits and error-tolerance parameters set to numerical accuracy. In our stochastic simulations, we determine the number of trajectories that are needed by analyzing the convergence of~\cref{eq: mcsolve convergence observable} for all the stabilizer and logical-qubit operators.
We found that 50k trajectories were sufficient to estimate these expectation values with a sampling uncertainty of less than $1\,\%$, for the reduced model that we introduce below.

\subsubsection{Measurement model}

Qubit measurements are modeled by letting the qubit idle for a time equal to the experimental readout time, followed by the projection of the stochastic wavefunction according to
\begin{equation}
    |\psi\rangle\to \frac{\Pi^j_k|\psi\rangle}{\sqrt{p^j_k}},
    \label{eq: projective measurement rule}
\end{equation}
where $\Pi^j_k=|j\rangle\langle j|_k$ is the single-qubit projector of the qubit~$k$, ~$j\in\{0,1\}$ corresponds to the measured qubit state, and $p^j_k=\langle\psi|\Pi^j_k|\psi\rangle$ is the probability of measuring qubit~$k$ in the state~$j$. We model qubit readout errors by flipping the result of the measurement with a probability that is computed from the readout assignment matrix for each qubit.

\subsubsection{Single-qubit gate model}

Single-qubit gates on qubit~$k$ are implemented with a Gaussian DRAG waveform with carrier frequency~$\omega_k$~\cite{Motzoi2009} and additional virtual-$Z$ gates~\cite{Mckay2017} when needed.
To speed up the simulations, we drop counter-rotating terms following the usual rotating-wave approximation (RWA).
We test this approximation by first comparing the average gate error per qubit with and without the counter-rotating terms, which we find to be limited by decoherence instead. Second, we have found no significant discrepancy in the estimated value of the logical qubit operators when using a RWA for simulating circuits with up to seven-qubits in a setup similar to that of Ref.~\cite{Andersen2020b} for up to five quantum-error-detection cycles.

\subsubsection{Two-qubit gate model}

\CZs{} between a pair of qubits~$(k,k')$ are emulated by the free evolution of the effective Hamiltonian~$(\pi/t_g-\xi_{k,k'})b^\dagger_k b_k b^\dagger_{k'}b_{k'}$, where~$t_g$ is the experimental gate time.
During the gate time, we change the dephasing rates~$\gamma_{\varphi,k}$ and~$\gamma_{\varphi,k'}$ of the two qubits involved in the gate to the effective values corresponding to the coherence times measured experimentally for the 24 different controlled-$Z$ gates at the qubit interaction frequencies, see \cref{app:two_qubit_gates}.

Given that the qubits are flux-biased at specific interaction frequencies during the gate time, we also account for the residual cross-Kerr interactions by adjusting the interaction strengths $\xi_{k,k'}$ of the gate qubits with their neighbors on the device. With this model, we simulate an average gate infidelity of $0.9\,\%$, which is consistent with the median of the interleaved randomized benchmarking error of $1.2\,\%$. Data post-selection in the experiment mitigates the impact of leakage to a large extent and enables the use of a simpler model of the two-qubit gates which involves levels only within the computational subspace. As discussed in~\cref{app:two_qubit_gates}, the performance of some of the two-qubit gates is limited by the interaction of qubits with strongly coupled two-level defects, an effect which is not included in the model used for the simulations.

\subsubsection{Pulse schedule}

Combining our model for the single- and two-qubit gates with projective measurements, we concatenate these operations to compose the pulse sequences used in the quantum error correction experiment, including buffer times, state-preparation and measurement pulses.

\subsection{Reduced 9+4 model}

By using the Monte Carlo solver we can simulate a 17-qubit system in a reasonable time using a general-purpose workstation: approximately 25~s per trajectory per error-correction cycle with a clock speed of~$\sim 3$~GHz and using about $6$~GB of RAM. However, improvements in runtime and potential extensions of this method to even larger systems are possible with effective models with a reduced number of qubits. In this section, we describe an approximate model that employs a total of 13 qubits and is obtained by tracing out the auxiliary-qubit modes that do not participate in a given stabilizer measurement.

\begin{figure*}[t] 
    \centering
    \includegraphics[width=1.\textwidth]{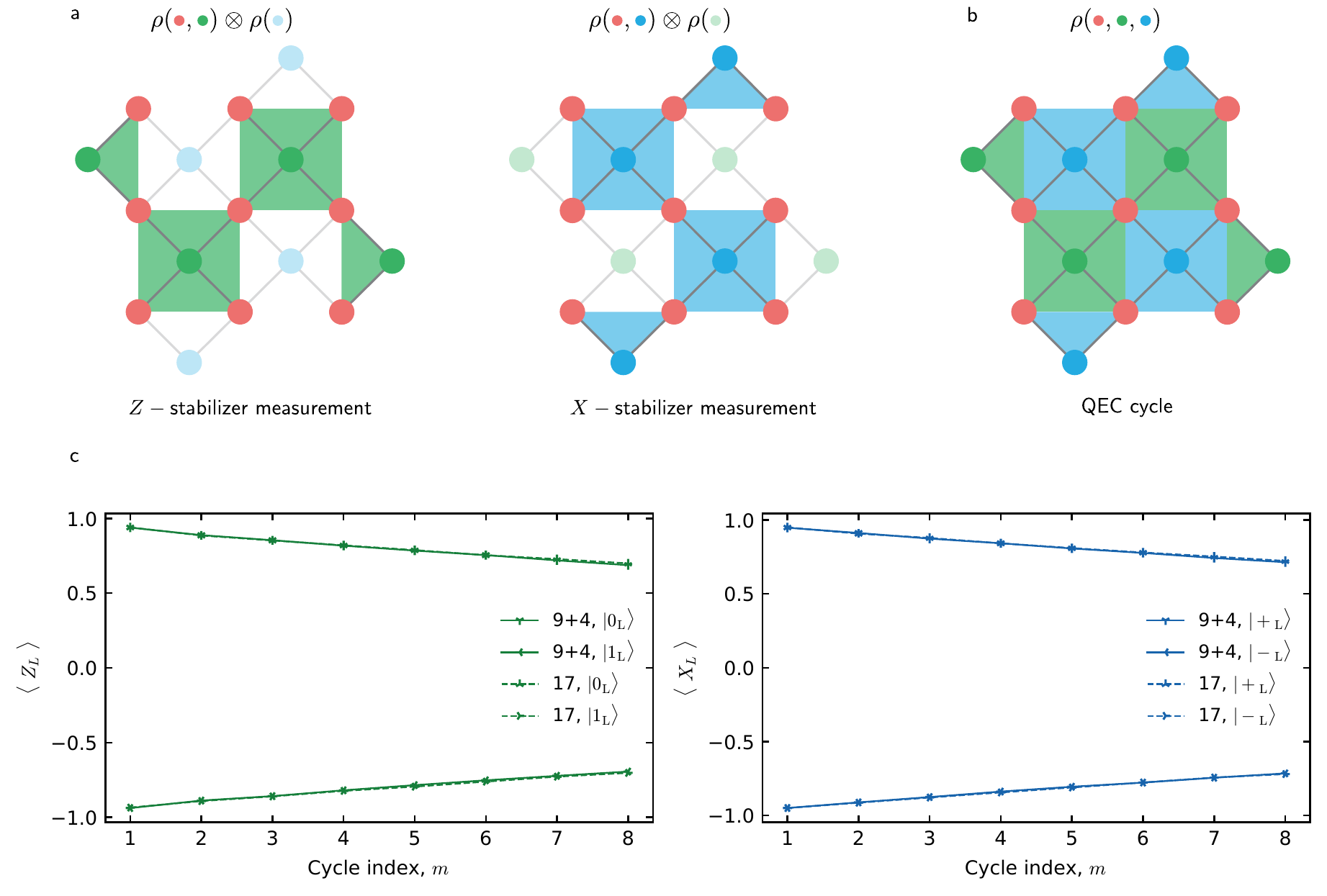}
    \caption{\textbf{a} Illustration of the 9+4 model. Logical~$Z$ ($X$) stabilizer measurements are simulated by tracing out the state of the~X-type (Z-type) auxiliary qubits prior to the stabilizer measurement.~$\rho(\{\bullet\})$ refers to a density matrix describing the subsystem~$\{\bullet\}$ of qubits indexed by color. \textbf{b} Full 17-qubit model includes all qubits at all times in the simulation.
    \textbf{c} Comparison between the decoded simulated data for the reduced 9+4- and the full 17-qubit model. The parameters of the simulation are those measured on the device and provided in~\cref{tab:qubit_measured_parameters}. We observe an
    excellent agreement of the resulting logical~$T_1$ and~$T_2$ times between these two models.}
    \label{fig: schematic 9 plus 4 model}
\end{figure*}

In practice, this effective model significantly reduces the simulation requirement to about 2~s per trajectory per QEC cycle while using less than $1$~GB of memory. A schematic illustration of the two models in consideration is provided in~\cref{fig: schematic 9 plus 4 model}.

To introduce the ``9+4'' model, we first consider the Z-type stabilizer measurements which constitute the first half of the QEC cycle. As illustrated in~\cref{fig: schematic 9 plus 4 model}a (see also~\cref{fig:stabilizer}a), the circuit only involves single- and two-qubit operations on the nine data and four Z-type auxiliary qubits. Meanwhile, the measurement of the X-type auxiliary qubits, which occurs in 400~ns, projects those qubits into a product state of the form~$|X_n\rangle=\bigotimes_{k''\in\mathrm{X-aux.}}|j_{k''}\rangle$ with~$j_{k''}\in\{0,1\}$. By assuming that the state of the X-type auxiliary qubits remains close to a product state during the full semi-cycle duration, it is possible to trace-out those qubit modes from the Hamiltonian. Following the measurement, the best description of the auxiliary modes is given by a product of single-qubit density matrices of the form~$|1_{k''}\rangle\langle 1_{k''}|e^{-t/T_{1,k''}} + |0_{k''}\rangle\langle 0_{k''}|(1-e^{-t/T_{1,k''}})$ if the measurement result for qubit~$k''$ is 1, and~$|0_{k''}\rangle 0_{k''}|$ otherwise. In other words, this corresponds to the effective model
\begin{align}
\begin{split}
    \hat{H}_\mathrm{red}(t)/\hbar &= \sum_{k} \Big(\omega_k +  \sum_{k''\in X} \xi_{k,k''}e^{-t/T_{1,k''}} \langle b_{k''}^\dagger b_{k''}\rangle\Big) b^\dagger_k b_k \\
    &+ \frac{\alpha_k}{2}b^{\dagger 2}_k b^2_k -i \Omega_k(t) (b_k - b^\dagger_k) \\
    &+ \sum_{k<k'} \xi_{k,k'} b^\dagger_k b_k b^\dagger_{k'}b_{k'},
    \label{eq: effective reduced circuit Hamiltonian}
\end{split}
\end{align}
where the subindices~$k$ and~$k'$ run over data and~Z-type auxiliary qubits, and~$\langle b_{k''}^\dagger b_{k''}\rangle$ is the expectation value of~$b_{k''}^\dagger b_{k''}$ immediately after the preceding~X-type auxiliary measurement. After evolving the 13-qubit wavefunction under $\hat{H}_\mathrm{red}(t)$ for the full semi-cycle, the projective measurement of the four Z-type auxiliary qubits is performed and the state of the system at time~$T$ is updated as
\begin{equation}
    |\psi(T)\rangle|Z_n\rangle\to|\psi(T)\rangle|X_n\rangle,
    \label{eq: state update}
\end{equation}
where $|\psi(T)\rangle$ describes the state of the data qubits after the projection. We perform the simulation of the following~X-type stabilizer measurement in a similar way.

In our simulations, we account for some of the noise associated with the auxiliary qubits that were traced out. First, as~\cref{eq: effective reduced circuit Hamiltonian} shows, the cross-Kerr interaction between an auxiliary qubit that has been traced out and its neighboring data qubits varies in time to account for the auxiliary qubit's finite~$T_1$ time. Second, the idling state of a traced-out auxiliary qubit~$k''$ after its measurement is modeled by flipping~$|j_{k''}\rangle$ from $|1\rangle$ to $|0\rangle$ with probability~$1-\exp{\left(-T_\mathrm{semi-cycle}/T_{1,k''}\right)}$, where~$T_\mathrm{semi-cycle}$ is the time of a quantum error correction semi-cycle. The resulting auxiliary-qubit state from this process is used to simulate the next quantum error correction cycle.

To test the validity of this reduced model, \cref{fig: schematic 9 plus 4 model}c shows a comparison between the results obtained for the reduced 9+4- and 17-qubit models for up to eight QEC cycles. The parameters of the simulation reflect those of the measured device, as discussed above. We show the decoded data from a logical-state preservation simulation as a function of the number of QEC cycles, from where we can infer the logical~$T_1$ and~$T_2$ for the two models. We use the decoder described in~\cref{app:decoding} with weights that are extracted from the simulation data for each model. The result shows an excellent agreement between the reduced and full models of the experiment, with a discrepancy that is contained in the error of the exponential fit. This observation gives us confidence about the accuracy of the 9+4 model used to perform the simulations reported in the main text. We note that, while limited in some respects, our effective model goes beyond previous approaches~\cite{OBrien2017,Huang2020} in the treatment of correlated errors (spurious~$ZZ$) and dissipation in continuous time.

\section{Stabilizers} \label{app:stabilizers}
The distance-three surface code comprises eight stabilizer operators, i.e.~products of $\hat Z$ or $\hat X$ Pauli operators of a subset of the nine data qubits, as listed in \cref{tab:stabilizers}.
We characterize the quantum circuit realizing each \sai{} (\cref{fig:stabilizer}a,b) for the $2^N$ input states of the $N \in \{2,4\}$ data qubits of a given plaquette (\cref{fig:device}a) and compute the stabilizer error
\begin{equation}
\epsilon = 1 - \frac{1}{2^N}\sum_{n=1}^{2^N} \frac{1}{2}|\overline{s}^{Ai}_n -\overline{s}^{Ai}_{n, \mathrm{ideal}}|,
\end{equation}
see \cref{tab:stabilizers}.
\begin{table}[]
    \centering
    \caption{Stabilizers \sai{} of the distance-3 surface code, characterized by the measured stabilizer error ($\epsilon$) and the stabilizer error ($\epsilon_\mathrm{sim}$) determined from simulation.}
    \begin{tabular}{clcc}
    \toprule
        Symbol & Stabilizer &  $\epsilon$ (\%) &  $\epsilon_\mathrm{sim}$ (\%) \\
        \midrule
         \sz{1} &  $\pz_1\pz_4$ & 2.9 & 2.8                     \\
         \sz{2} &  $\pz_4\pz_5\pz_7\pz_8$ & 8.4 & 5.0                   \\
         \sz{3} &  $\pz_2\pz_3\pz_5\pz_6$ & 6.8 & 4.3                   \\
         \sz{4} &  $\pz_6\pz_9$ & 2.5 & 2.0                     \\
         \midrule
         \sx{1} &  $\px_2\px_3$ &  5.7&  6.7                    \\
         \sx{2} &  $\px_1\px_2\px_4\px_5$ & 5.9 & 3.9                   \\
         \sx{3} &  $\px_5\px_6\px_8\px_9$ & 11.8 & 4.4                   \\
         \sx{4} &  $\px_7\px_8$ & 4.5 & 2.6                     \\
\midrule
	  & Weight-two average & 3.9 & 3.5 \\
	  & Weight-four average & 8.2 & 4.4 \\
	  & Average & 6.1 & 3.9 \\
\bottomrule
    \end{tabular}
    \label{tab:stabilizers}
\end{table}
We measure an average weight-two stabilizer error of $3.9\,\%$, which is in good agreement with the $3.5\,\%$ average weight-two stabilizer error extracted from numerical simulations (\cref{app:numerical_simulations}).
The average weight-four stabilizer error obtained from measurements ($8.2\,\%$) is larger than the one obtained from simulation ($4.4\,\%$), which we identify as most likely due to the interaction with microscopic defects, residual coherent errors in two-qubit gates, and residual crosstalk which are not modeled in our numerical simulations (\cref{app:numerical_simulations}).

\section{Logical state initialization and characterization} \label{app:logical_state_fidelity}
Here, we describe the characterization of the  prepared nine-data-qubit logical states by measuring their quantum state fidelity  with respect to (i) the target state, (ii) the target logical subspace, and (iii) states in correctable subspaces.

\subsection{Fidelity with respect to the target state}
While the complete tomographic reconstruction of a generic quantum state $\rho$ of $n=9$ qubits would require the measurement of $4^n = 262144$ independent Pauli correlators ~\cite{Nielsen2000}, the measurement of the fidelity with respect to a target logical state $ \rho_{\lz} =|0\rangle_{\mathrm{L}}\langle 0|_{\mathrm{L}} = $\prlogicalsubspace{}\prlogicaloperator{}
expressed in terms of the projector \prlogicalsubspace{} onto the logical subspace and the projector \prlogicaloperator{}  onto an eigenstate of $Z_L$
\begin{align}
    \prlogicalsubspace{}\prlogicaloperator{}
    & = \frac{1}{2^9} (1 + \lopz) \prod_{i = 1}^{4} (1 + s^{\mathrm{X}i}\hat S^{\mathrm{X}i}) \prod_{i = 1}^{4} (1 + \hat S^{\mathrm{Z}i}) \label{eq:rho_z_prod}\\
    &= \frac{1}{512}\sum_{j=1}^{512} \gamma_j \hat P_j \label{eq:lz_pauli}
\end{align}
requires the measurement of only $2^n =512$ terms ~\cite{Nigg2014b, Abobeih2021}. Here, $s^{\mathrm{X}i}$ are the outcomes of the $X$-stabilizer measurements obtained in the state initialization,  $\hat P_i$ are 9-qubit Pauli correlators obtained from expanding the product in \cref{eq:rho_z_prod}, and the $\gamma_j$ take values $+1$ or $-1$ depending on the individual outcomes of $\{s^{\mathrm{X}i}\}$.
The fidelity $ \mathcal{F_\mathrm{phys}}$ of state $\rho$ with respect to \lz{} is given by
\begin{equation}
        \mathcal{F_\mathrm{phys}}  = \trace \left( \rho \rho_{\lz} \right)
        = \frac{1}{512} \sum_{j=1}^{512}\langle \gamma_j \hat P_j \rangle_\rho \label{eq:fid_phys_pauli}
\end{equation}
where $\langle  \gamma_j \hat P_j \rangle_\rho$ corresponds to the expectation value of the Pauli correlator $\hat P_i$ in state $\rho$.

We evaluate each of the 512 Pauli correlators of \cref{eq:fid_phys_pauli} with data collected by executing a single quantum error correction cycle, followed by single-qubit tomography rotations on data qubits and a readout of all qubits.
We observe that in ~80.5\% of the experimental runs the measurements of $s^{Zi}$ yield the target value +1 for all four \sz{i}, which is in good agreement with the probability of having none of the four Z-type mean syndrome elements signaling an error in the first cycle,
$\prod_i^4(1-\overline{\sigma}_1^{Zi}) \approx 84.5\%$, see  \cref{fig:stabilizers_state_init}c.

For each $\langle\gamma_j \hat P_j \rangle_\rho$, we reject leakage events as detected by our three-state readout scheme (\cref{app:leakage_rejection}) and account for readout errors (\cref{app:readout_characterization}) on each of the data qubits involved in the correlator. Note that apart from leakage rejection we keep all instances for further data analysis, including those events in which at least one of the four measured \sz{i} values yields $-1$.
Following \cref{eq:fid_phys_pauli}, we then take the average over all expectation values to compute $\mathcal{F_\mathrm{phys}}$.

\subsection{Fidelity in the logical subspace}
Similarly, we evaluate the probability of having prepared a state in the logical subspace as
\begin{equation}
P_\mathrm{L} = \trace \left( \rho \prlogicalsubspace\right) = \frac{1}{256} \sum_{j=1}^{256}\langle \gamma_j \hat P_j \rangle_\rho
\end{equation}
where the sum extends over the expectation values of the 256 Pauli correlators resulting from the expansion of the projector onto the logical subspace \prlogicalsubspace. Together with the value of $ \mathcal{F_\mathrm{phys}}$ the probability $P_\mathrm{L}$ yields an estimate for the logical fidelity $ \mathcal{F_\mathrm{L}} =  \mathcal{F_\mathrm{phys}}/P_\mathrm{L}$, which corresponds to the fidelity of the prepared state with the target logical state conditioned on having successfully projected onto the logical subspace.

\subsection{Fidelity with respect to states in correctable subspaces}
In the context of quantum error correction, it is insightful to compute the fidelity of the prepared state with respect to any of the states which are equivalent to the target state up to a correctable error.
By construction, a data-qubit state projected to the logical $Z$-basis is a $\pm 1$ eigenstate of the eight stabilizers $\sai$ and a $\pm 1$ eigenstate of $\lopz{}$. Thus, there are  $2^9 = 512$ distinct eigenstates of the data-qubit space in the logical Z-basis, with pair-wise degenerate syndromes.
However, in a simple error correction scheme, each of the $2^8 = 256$ possible syndromes produced by the eight stabilizers is associated to a single corrective action. Consequently, for each of the 256 pairs of eigenstates with a degenerate syndrome, only one of the two eigenstates can be included in a given correctable subspace. To perform error correction most effectively, one  includes the eigenstate in the correctable subspace, which can be reached by a Pauli error that is more likely to occur on the physical device.

Specifically, we construct correctable subspaces for \lz{} in the following way: for each pair of eigenstates with degenerate syndrome, we compute the lowest weight errors which, when applied to \lz{}, lead to these two eigenstates, respectively. With the assumption that lower-weight errors are more likely than higher-weight errors, we preferably include the eigenstate that can be reached via the lowest weight error.
If the two eigenstates can only be reached by applying errors of equal weight-$n$, we check how many different errors of weight-$n$ lead to each of the two eigenstates, and include the eigenstate that can be reached by more weight-$n$ errors.
Finally, if both eigenstates can be reached by an equal number of weight $n$ errors, we randomly select one of the two eigenstates.
Note that with this approach, any correctable subspace includes $\lz{}$ (trivial +1 eigenstate of all stabilizers and of \lopz), as well as all the states, which  can be reached by applying a weight-one Pauli error on any one of the data qubits.

We define the fidelity of the prepared state $\rho$ with respect to a set of $N$ states $\{\hat E_n\lz{}\}$ as
\begin{align*}
        \fcorrectable & = \sum_{n=1}^N \trace \left( \rho \, \hat E_n|0\rangle\langle 0|_{\mathrm{L}}\hat E_n^\dagger \right) \\
							  & = \frac{1}{512} \sum_{n=1}^N \trace \left( \rho \sum_{j=1}^{512} \hat E_n \gamma_j \hat P_j \hat E_n^\dagger \right)
\end{align*}
where we have made use of \cref{eq:lz_pauli}.
We observe that $\hat E_n  \hat P_j \hat E_n^\dagger = c_{j, n} \hat P_j$ with $c_{j, n} = \pm 1$ because  $\hat E_n$ and $\hat P_j$ are both tensors of Pauli matrices, and all matrices in the Pauli group commute or anti-commute.
Consequently,
\begin{equation}
 \fcorrectable = \frac{1}{512} \sum_{n=1}^N \sum_{j=1}^{512}  c_{j, n} \langle \gamma_j \hat P_j \rangle_\rho  \label{eq:fid_corr_pauli}
\end{equation}
can be computed based on the same set of 512 measured expectation values $\langle \gamma_j \hat P_j \rangle_\rho $ with the appropriate combination of sign prefactors $c_{j, n}$. This circumvents the need to measure the expectation value of 512 Pauli correlators for each of the $N=256$ states spanning the correctable subspace.

We evaluate the fidelity \fcorrectable{} of the prepared state with respect to any state in the correctable subspace for 500 different correctable subspaces randomly chosen according to the procedure described above. We obtain a mean correctable fidelity averaged over the 500 correctable subspaces of $\fcorrectable = 96.0(9)\, \%$.
The 4.0(9)\% average infidelity, which can also be interpreted as the probability of being in a state which would result in a logical error after a single cycle, is in good agreement with the probability of logical error per cycle $\epsilon_{\rm L}=0.032(1)$ deduced from the state preservation of $\langle\lopz\rangle$, see main text. We observe that $\epsilon_{\rm L}$ is slightly smaller, likely due to the fact that the state initialization characterization cannot correct for auxiliary-qubit readout errors, which require several quantum error-correction cycles to be detected.

\section{Pulse sequence} \label{app:pulse_sequence}
\begin{figure*}[!t] 
\centering
\includegraphics[width=\textwidth]{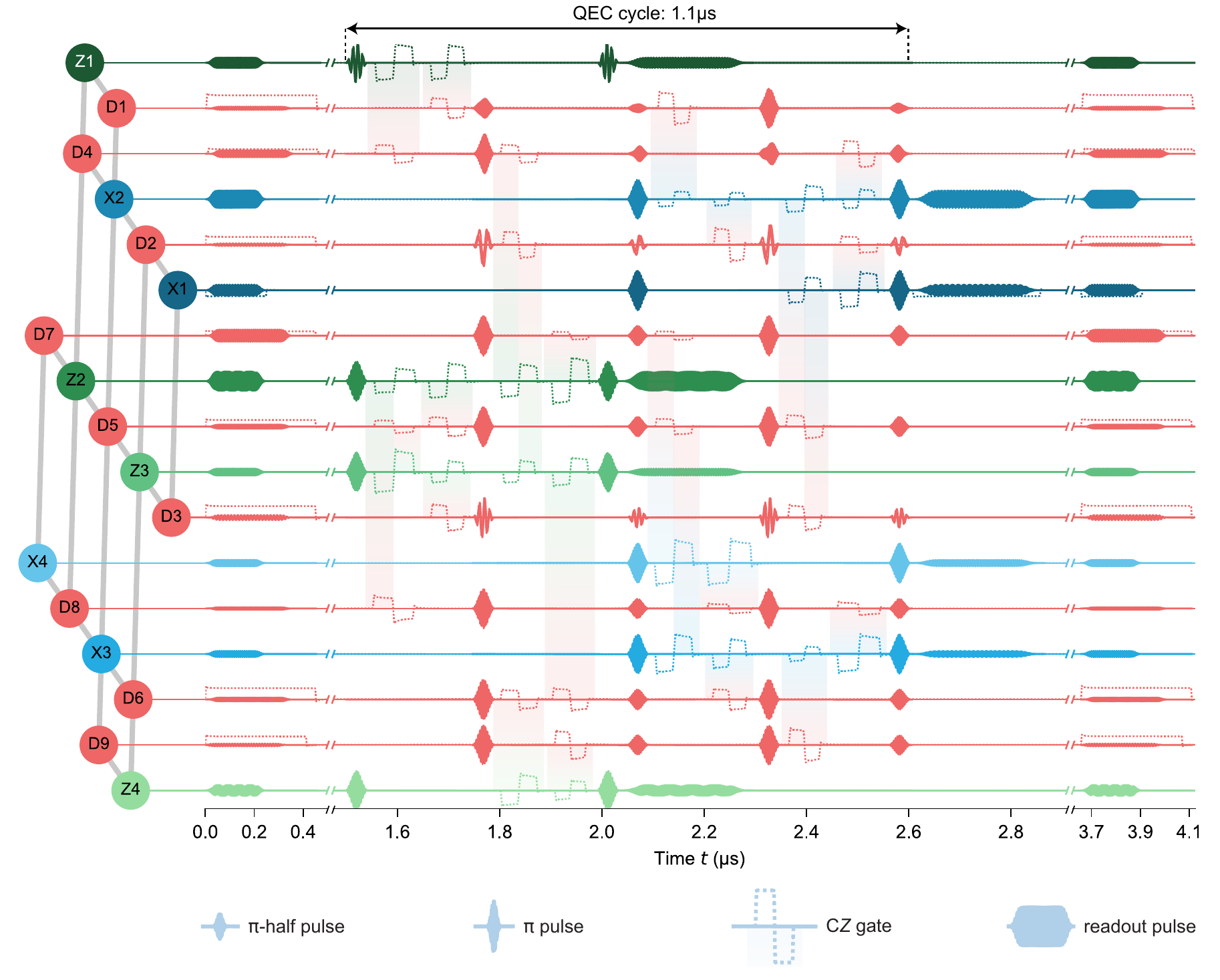}
\caption{AWG pulse sequence for the execution of a single quantum error correction cycle. Single-qubit drive pulses (solid lines), readout pulses (solid lines), and flux pulses (dotted lines) are displayed for each of the 17 qubits, as indicated by the label in the conceptual device representation. Qubit pairs realizing \CZs{} are connected by shaded gradients. The time axis is segmented (parallel lines intersecting the axes) during idle times for clarity, see text for details.  }
\label{fig:pulse_sequence}
\end{figure*}

We realize the quantum circuit presented in \cref{fig:stabilizers_state_init}\,a with a combination of microwave and flux pulses applied to the 17 qubits, see \cref{fig:pulse_sequence} for the complete pulse sequence of a single cycle of quantum error correction used to prepare \lz{}.

The sequence starts with a multiplexed readout of all qubits to herald the ground-state in post-selection. Next, single-qubit drive pulses are applied to a subset of the data qubits to prepare the nine data qubits in one of the four product states $\ket{0}^{\otimes 9}$,  $\lopx{}\ket{0}^{\otimes 9}$, $\ket{+}^{\otimes 9}$ or $\lopz{}\ket{+}^{\otimes 9}$, each of which is a superposition of the $16$ equivalent instances of the respective target logical state.
Thereafter,  Z-type stabilizers are executed by applying three simultaneous \CZs{} in four sequential time steps, with a dynamical decoupling pulse applied to all data qubits in between the second and the third two-qubit gate time step. Each \CZ{} is realized by applying a flux pulse to a data qubit and to the corresponding auxiliary qubit, see the shaded gradient connecting pairs of qubits in \cref{fig:pulse_sequence}.
While the Z-type auxiliary qubits are being read out, the X-type stabilizers are realized in a similar fashion as the Z-type stabilizers but with $\pi/2$-rotations implementing basis changes before and after the four time steps of \CZs. Finally, all qubits are read out in the Z-basis after a latency of \ns{800} set by the minimum re-triggering period of our acquisition devices.

\section{Leakage rejection} \label{app:leakage_rejection}

While non-computational states can provide a useful resource for example to realize two-qubit gates in transmon qubits,
uncontrolled leakage into non-computational states constitutes a source of errors.
Leakage has therefore been addressed both theoretically and experimentally by reducing leakage errors \cite{Chen2016, Negirneac2021}, by detecting leakage indirectly on both auxiliary and data qubits using hidden Markov models \cite{Bultink2020} or quantum non-demolition measurement protocols \cite{Stricker2020}, by developing leakage-aware decoding schemes \cite{Suchara2015, Kelly2015}, and by converting leakage errors into errors within the computational subspace rendering them correctable in the standard error correction framework \cite{Alferis2007, Fowler2013, Ghosh2015a}.

For our experimental realization of the surface code we mitigate leakage, particularly of data qubits, by choosing a frequency configuration and two-qubit gate scheme in which only the auxiliary qubits evolve through the second excited state $|2\rangle$, keeping the average leakage probability per data qubit and per cycle as low as about two per mill. Furthermore, we detect the remaining leakage events on both the data and auxiliary qubits by implementing a high-fidelity three-state readout scheme (\cref{app:readout_characterization}), which allows us to reject all instances of experimental runs with detected leakage events and to study the performance of quantum error correction independent of leakage dynamics \cite{Varbanov2020}.

\begin{figure}
\centering
\includegraphics{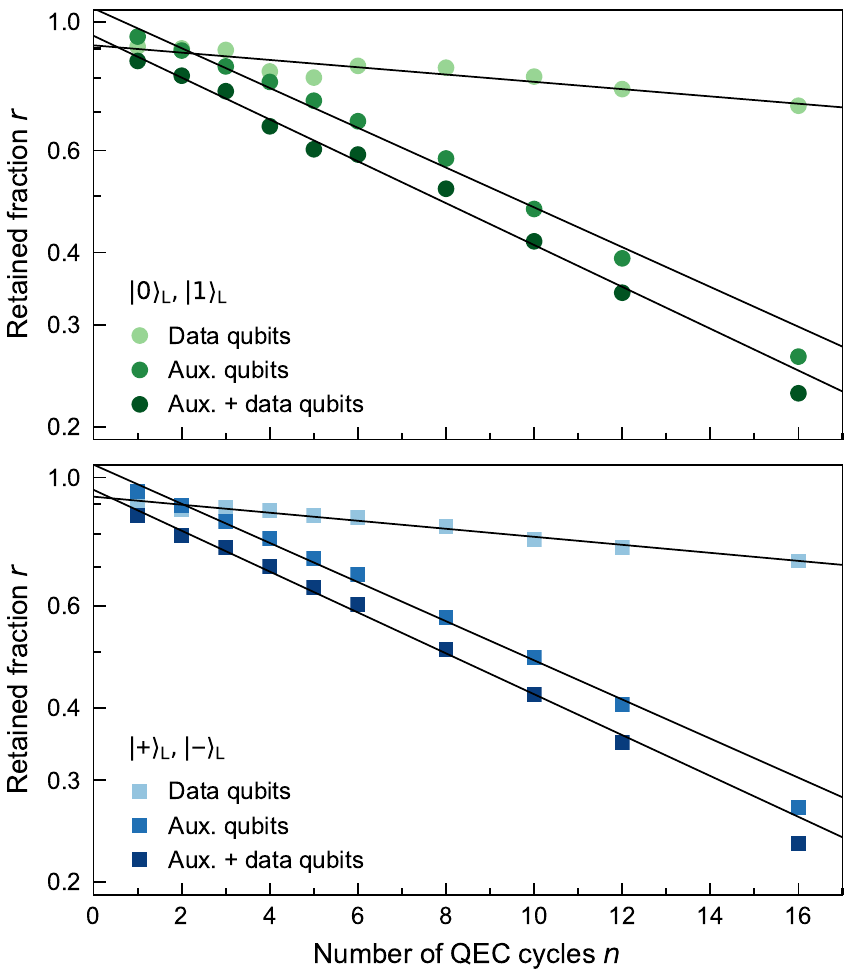}
\caption{Retained fraction of experimental runs after leakage rejection. \textbf{a} Retained fraction when rejecting leakage of data qubits only (light, green symbols), of auxiliary qubits only (mid-green), and of both (dark green). Solid black lines are exponential fits to the data. Data corresponds to the state preservation experiments of \lz{} and \lo{} presented in the main text. \textbf{b} Same as \textbf{a}, but for data from the state preservation experiments of \lp{} and \lm{}.}
\label{fig:retained_fractions}
\end{figure}

When analyzing the state preservation experiments shown in Fig.~\ref{fig:state_preservation} of the main text, we find that the retained fraction of runs $r$ in which no leakage event has been detected, decreases to good approximation exponentially with $n$, indicating that the retained fraction of runs per cycle $r_{\rm c}$
is independent of the cycle number, see Fig.~\ref{fig:retained_fractions}. The value $r_{\rm c}\approx 92.1(3)\,\%$, which we obtain from exponential fits $r=Ar_{\rm c}^n$ (black lines) to the data (dark symbols) is identical within error bars for prepared eigenstates of $\hat Z_{\rm L}$ (panel a) and $\hat X_{\rm L}$ (panel b). We attribute the slight deviation between the measured data and the exponential fit to the finite probability of falsely classifying a $|1\rangle$-state as $|2\rangle$ during readout, which for auxiliary qubits is on average $0.004$ and for data qubits $0.019$. In fact, the probability for having such falsely identified leakage events is proportional to the population of the $|1\rangle$-state, which for auxiliary qubits increases from cycle to cycle due to their initialization in $|0\rangle$ and the error correction protocol, approaching 0.5 in the limit of large $n$. Hence, the falsely identified auxiliary-qubit leakage events increase with $n$, approaching about $21\,\%$ of the totally detected auxiliary qubit leakage events and explaining the slight increase of detected leakage events per cycle with $n$.

Distinguishing between leakage events on data and auxiliary qubits (lighter data points in \cref{fig:retained_fractions}\,a,\,b), we find that the probability for having detected a leakage event on any of the nine data qubits is only about $0.015(2)$ per cycle and about $0.075(3)$ for the eight auxiliary qubits.

For comparison, we also determine the extracted logical error probabilities $\epsilon_{\rm L}$ without leakage rejection (i), when rejecting data qubit leakage only (ii), and when rejecting auxiliary qubit leakage only (iii), see Table~\ref{tab:leakage_schemes}. For simplicity, we interpret non-rejected measured $|2\rangle$-states as $|1\rangle$-states during decoding. Compared to the case when rejecting all detected leakage events (iv), the average absolute increase in the logical error probability is $0.021(2)$ for scheme (i), $0.016(1)$ for scheme (ii), and only $0.001(1)$ for scheme (iii).
These results suggest that the low leakage rates achieved on our device, when combined with auxiliary-qubit reset using either feedback in combination with three-state readout or an unconditional scheme \cite{Magnard2018, McEwen2021a}, could render leakage errors tractable in general.

\begin{table}[]

    \centering
        \caption{Extracted logical error per cycle $\epsilon_{\mathrm{L}}$ for the experiment preserving eigenstates of \lopz{} and of \lopx{} using the indicated leakage rejection schemes. The retained data fraction per cycle $r_c$ after leakage rejection is also indicated.}
    \label{tab:leakage_schemes}
    \begin{tabular}{lrrc}
    \toprule
        Leakage rejection & $\epsilon_{\mathrm{L}}\, [\lopz]$ & $\epsilon_{\mathrm{L}}\, [\lopx]$ & $r_c$ \\
        \midrule
        (i)\,\, None & 0.054(1) & 0.049(2) &  1.000(0)              \\
        (ii)\, Data qubits only & 0.049(1) & 0.044(1)  &  0.985(2)               \\
        (iii) Aux. qubits only & 0.033(2) & 0.030(1) &  0.925(3)                 \\
        (iv) Aux. and data qubits & 0.032(1) & 0.029(1) &  0.921(3)   \\
    \bottomrule
    \end{tabular}

\end{table}

\section{Decoding and weight extraction}\label{app:decoding}
To identify the most likely sequence of errors having occurred in a single instance of an experimental run, we decode the measured set of syndromes in post-processing by adopting the minimum-weight-perfect matching (MWPM) algorithm described in Ref.~\cite{OBrien2017}. We represent each syndrome element $\sigma_{m}^{Ai}$ as a vertex of a graph, which has two spatial dimensions set by the layout of auxiliary qubits ${Ai}$ in the surface code lattice, and one temporal dimension indexed by the cycle number $m$. We connect pairs of vertices with edges, each of which represents a particular error at the physical level \cite{Kelly2015}.

In our model, we consider three kinds of errors. First, auxiliary qubit errors, including errors during readout, which are represented by vertices at the same location but separated in time by one cycle ($\Delta m =1$). Second, auxiliary qubit measurement misclassification errors, i.e. measurement errors which do not change the state of the auxiliary qubit, are represented by edges connecting vertices at the same location but separated in time by two cycles ($\Delta m =2$). And third, single Pauli errors acting on data qubits, which are represented by edges connecting vertices either to a direct neighbor of the same type ($X$ or $Z$) or to a boundary. Depending on the time at which these errors occur, the corresponding edge either connects to a vertex of the same cycle ($\Delta m =0$) or the next cycle ($\Delta m =1$).   We extract the probabilities associated with those edges directly from measured syndrome correlations using the methods described in Refs.~\cite{Spitz2018, Chen2021p}. Details of this scheme will be provided in a separate publication.

Based on the individual edge error probabilities, we then compute for \emph{all} pairs of vertices $k$ and $l$ with $\Delta m \leq 2$, the total probability $p_{{\rm }kl}$ of being connected. We do so, by summing the individual probabilities over all possible error paths along the edges of the graph, see Ref.~\cite{Spitz2018} for details. Here, the terms  $p_{{\rm }kk}$ correspond to the probability of having vertex $k$ being connected to a boundary.  We convert the matrix of probabilities ${p_{{}kl}}$ into a weight matrix $w_{kl} = -\ln{p_{{}kl}}$, based on which the MWPM algorithm connects each syndrome element $\sigma_{m}^{Ai}=1$ either to a second such syndrome element or to a boundary while minimizing the total weight associated with these connections.

To correct for decoded errors in the final outcome $z_{\rm L}$ ($x_{\rm L}$) of the logical operator $\hat{Z}_{\rm L}$ ($\hat{X}_{\rm L}$), we evaluate ${z}_{\rm L}=z_1 z_2 z_3$ ($x_{\rm L}=x_1 x_4 x_7$) from the final data qubit readout and multiply it by $(-1)^M$, where $M$ is the number of syndrome element pairs determined by the MPWM algorithm, which signal a logical error. A syndrome element pair signals a logical error if the underlying error path contains an odd number of errors on those data qubits, which are contained in the logical operator string $\hat{Z}_{\rm L} = \hat{Z}_{1}\hat{Z}_{2}\hat{Z}_{3}$ ($\hat{X}_{\rm L} = \hat{X}_{1}\hat{X}_{4}\hat{X}_{7}$).

For consistency we evaluate the weights for each of the experimental and simulated state preservation experiments independently. We verified that the decoding with weights extracted from separate data sets yields the same logical lifetimes.

\end{appendix}


\bibliography{resources/QudevRefDB,resources/references}

\end{document}